\documentclass[12pt]{article}
\usepackage{authblk}
\usepackage{graphicx}
\usepackage{framed}
\usepackage{amsfonts}
\usepackage{amsmath}
\usepackage{array}
\usepackage{verbatim}
\usepackage{psfrag}
\usepackage{subcaption}
\usepackage{tikz}
\usetikzlibrary{arrows.meta}
\usepackage[makeroom]{cancel}
\usepackage{circuitikz}
\usepackage{wrapfig}
\usepackage{multirow}
\usepackage{rotating}
\usepackage{soul}
\usepackage{color}
\usepackage{appendix}
\usepackage{sansmath}
\usetikzlibrary{shadings,intersections}

\begin{document}

\title{Comparing the Electric Fields of Transcranial Electric and Magnetic Perturbation}


\author[1]{D. J. Sheltraw}
\author[1]{B. Inglis}
\author[2]{L. Labruna}
\author[2]{R. Ivry}
\affil[1]{Henry H. Wheeler, Jr. Brain Imaging Center, University of California Berkeley}
\affil[2]{Department of Psychology and Helen Wills Neuroscience Institute, University of California Berkeley}

\date{\today}

\maketitle

\begin{abstract}

Noninvasive brain stimulation (NIBS) by quasistatic electromagnetic means is presently comprised of two methods: Magnetic 
induction methods (Transcranial magnetic perturbation or TMP) and electrical contact methods (Transcranial electric perturbation 
or TEP). Both methods couple to neuronal systems by means of the electric fields they produce. Both methods are necessarily 
accompanied by a scalp electric field which is of greater magnitude than anywhere within the brain. A scalp electric field of 
sufficient magnitude may produce deleterious effects including peripheral nerve stimulation and heating which consequently limit 
the spatial and temporal characteristics of the brain electric field. Presently the electromagnetic NIBS literature has produced 
an accurate but non-generalized understanding of the differences between the TEP and TMP methods. The aim of this work is to 
contribute a generalized understanding of the differences between the two methods which may open doors to novel TEP or TMP methods 
and translating advances, when possible, between the two methods. This article employs a three shell spherical conductor head model
to calculate general analytical results showing the relationship between the spatial scale of the brain electric fields and: (1) 
the scalp-to-brain mean-squared electric field ratio for the two methods and (2) TEP-to-TMP scalp mean-squared electric field ratio
for similar electric fields at depth. The most general result given is an asymptotic limit to the TEP-to-TMP ratio of scalp 
mean-squared electric fields for similar electric fields at depth. Specific example calculations for these ratios are also given 
for typical TEP electrode and TMP coil configurations. While TMP has favorable mean-squared electric field ratios compared to TEP 
this advantage comes at an energetic cost which is briefly elucidated in this work.

\end{abstract}

\section{Introduction}

The neuronal tissue of the brain can be perturbed noninvasively by the application of an electric
field generated by two means: Magnetic induction and electrical contact. The magnetic induction 
method uses a time varying current within coils external to the head, and not in electrical contact 
with the head, to produce a time varying magnetic field within the brain. This time varying magnetic
field induces an electric field in the electrically conductive head. The electric contact method 
uses a source of current and electrodes in contact with the head to produce an electric field 
within the brain. Regardless of which method is used, the electric field is stronger in the scalp than in the brain. Therefore, when designing new systems to perturb brain function, it is of considerable importance to understand with some generality the characteristics of the brain and scalp electric fields of each method. In addition it is important for the researcher to understand the energetic costs of generating electric fields by each method.

Here the electric contact method will be referred to as Transcranial Electric Perturbation (TEP) 
and the magnetic induction method will be referred to as Transcranial Magnetic Perturbation (TMP). 
Therefore TMS (transcranial magnetic stimulation), which is an induction method that employs brief
(approximately 250 $\mu$s) and possibly intense pulses (requiring as much as 6 kA of coil current)
capable of producing suprathreshold electric fields (greater than 40 V/m) within the brain, is a 
TMP method. Similarly TES (transcranial electric stimulation), which is a electric contact method 
that employs sustained (300-1800 s) subthreshold electric fields (approximately 0.5 V/m) generated 
from relatively small contact currents (typically up to 2.0 mA in the DC--kHz range) or 
electroconvulsive therapy (ECT), which is an electric contact method that employs brief pulses (0.2--2.0 ms) delivered as a train of brief pulses (0.2--1.0 ms square wave pulse, 20--240 pulses/s, $\le$ 8 s total duration) 
using contact current amplitudes in the range of 100--900 mA, are both TEP methods.  

Regardless of the method used the resulting electric field couples to neurons and may perturb their state in a short or long term manner. When the electric field is suprathreshold, robust effects 
such as spiking and electrical nerve blocking \cite{KilBha1} can be elicited with kHz continuous waveforms. When the electric 
field is subthreshold effects such as entrainment \cite{Deans_1}\cite{Jeff_1} and motor threshold changes \cite{Mikkonen_1}\cite{Chaieb} can be elicited. The dynamics of the coupling of the applied electric field to any given neuron can be described by equations that predict the change in the neuron's transmembrane potential (the ouput) which are in general nonlinear with respect to the
applied electric field (the input) \cite{Gabb}. If the electric field amplitude is much smaller 
than threshold then the electrodynamics of the neuronal system can often be described by a linear relationship between the input electric field and the output transmembrane potential. As the input amplitude increases the linear approximations will fail and nonlinear relationships must 
ultimately be employed.

Most TEP and TMP modeling employs the finite element method (FEM) in conjunction with volume conductor models built from magnetic resonance images \cite{Opitz}\cite{Datta_1}\cite{Datta_2}\cite{Datta_3}\cite{Mendonca_1} to estimate the electric field within the head. However such detail is not necessarily needed or even desirable when trying to establish general physical and engineering principles associated with the TEP and TMP methods. In fact, when making comparisons between these methods, numerical calculation of electric fields generated by specific TEP electrode geometries or specific TMP coil geometries can miss general principles, like those described in the body of this paper, which are obtainable through analytical calculations. 

The work herein makes clear, in a general manner, that the choice between the TEP or TMP methods depends primarily 
upon the temporal and spatial characteristics of the desired electric field as well as the energy consumption of 
the respective current sources. With respect to the electric field spatial characteristics it is shown that the 
TEP and TMP methods differ fundamentally with respect to the electric field subspaces they span and the 
scalp-to-brain power dissipation ratios they produce. All other differences, such as field focality, follow from 
these two general differences. 

Throughout this paper comparisons will be made between TEP and TMP electric fields using average quantities within the brain and scalp regions. In a given head region the natural single number proxies for the electric field magnitude and the absorbed power are the root-mean-squared and the mean-squared electric field respectively. These quantities are of great experimental consequence since the amplitude of the scalp electric 
field may limit the safely obtainable amplitude of the cortical electric field. Indeed, scalp peripheral nerve 
stimulation (which can range from distracting to painful) scales with root-mean-square electric field amplitude while scalp heating (which can range from benign to burning) scales with mean-square electric field amplitude. Note that since the conductivities of the brain and scalp regions are comparable and often assumed to be equal, as is often the case in three-shell models, then the scalp-to-brain power dissipation ratio is equivalent to the scalp-to-brain mean-squared electric field ratio. Also note that most extant quasistatic EM NIBS methods are limited by peripheral nerve stimulation rather than tissue heating. However, this may not apply to future methods in which electric field amplitude, frequency (although still quasistatic) and duration of the perturbing waveforms could be increased.

To present the differences between TEP and TMP electric fields in a clear manner a three-shell head model is 
employed and solved analytically. In this model the head is assumed to consist of three concentric 
spherically symmetric regions of differing conductivities which adequately represent the electromagnetic 
properties of the scalp, skull and brain. Vector spherical harmonics \cite{VMK} are used to 
describe the TEP and TMP electromagnetic fields and sources of current. This is a natural choice for the 
vector fields given the spherical geometry of the model. 

Most of the earlier treatments of the electric field within a spherically symmetric conductor did not make 
use of vector spherical harmonics and as a result the derivations were somewhat long and cumbersome \cite{HEaton}. 
The authors know of only two publications \cite{KNI} \cite{WangPeterchev}(articles concerned with TMS coil design) which make 
use of vector spherical harmonics in the treatment of such problems. However, scalar spherical harmonics have
been used in the analytic solution of the three-shell TEP model \cite{Stecker_1} albeit with skull, cerebrospinal 
fluid and brain as the three compartments of the model. That publication noted that the results of their 
calculations were only slightly dependent upon the conductivity and thickness of the CSF hence that compartment 
is not included in the present work. Here, for the first time, vector spherical harmonics are used 
to describe both the TMP and TEP electric fields thereby allowing for a direct comparison of the respective 
electric fields and properties. A real head will of course not be spherically symmetric nor will it be precisely 
separable into only three regions of differing electric conductivity, however the general principles and 
estimates established in this work apply approximately to more realistic models as well. 

This paper is organized as follows: In section \ref{theory} the three-shell model is solved for the electric 
field in the three regions modeling the scalp, skull and brain. Briefly the spatial differences between the TEP
and TMP electric fields are mentioned. In section \ref{power} the quantities $R^{tep}$, $R^{tmp}$ and $R$ are 
calculated. The quantities $R^{tep}$ and $R^{tmp}$ are the scalp-to-brain ratios of power dissipation for the 
TEP and TMP cases respectively whereas $R$ is the TEP-to-TMP ratio of scalp energy dissipation for the case of 
similar TEP and TMP electric fields at the radial position of the cortex. Example calculations of each ratio are 
given for the case of typical electrode and coil geometries. The energetic cost of generating an electric field 
within the brain depends upon the method used. Therefore section \ref{energy} presents a simple analysis of the 
power utilization of TEP and TMP current sources. In this manner a more complete picture of the benefits and 
costs of each method can be understood. The paper ends with a discussion of future methods that could potentially 
take advantage of the benefits of TMP albeit at a cost in power utilization and requiring new designs for TMP coil 
cooling systems.

\section{Methods}

\label{theory}

\begin{figure}[ht]
\centering
\tikzstyle{myPath} = [-{Latex[length=4mm,width=4mm]}]
\begin{tikzpicture}[scale=0.6]
  \draw (0,0) circle (6);
  \draw (0,0) circle (5);
  \draw (0,0) circle (4);
  \draw [myPath] (0,0) -- (2.5,4.33);
  \draw [myPath] (0,0) -- (4.0,-0.2);
  \draw [myPath] (0,0) -- (5.196,3);
  \node at (2.2,-0.5) {$r_0$}; 
  \node at (0.8,2.2) {$r_1$}; 
  \node at (2.2,0.75) {$r_2$}; 
  \node at (0,3.4) {$\sigma_0$};
  \node at (0,4.4) {$\sigma_1$}; 
  \node at (0,5.5) {$\sigma_2$}; 
  \node at (0,-3.3) {region 0}; 
  \node at (0,-4.4) {region 1}; 
  \node at (0,-5.5) {region 2}; 
\end{tikzpicture}
\caption{The three-shell spherical head model in which the regions from outermost to innermost are the scalp, 
        skull and brain respectively. Typical estimates for three shell model radii are $r_0 =$ 80 mm, 
        $r_1 =$ 86 mm and $r_2 =$ 92 mm whereas typical estimates $\sigma_0 = \sigma_2$ and $\epsilon = \sigma_1/\sigma_0 = 1/80$. Note that only the ratio $\epsilon$, rather than the specific values of the conductivities, is of importance in this work since the focus is on the calculation of mean-squared electric field ratios between the brain and scalp regions. The mean-squared electric field is a single number proxy for the energy dissipation in a region whereas it's square root is a single number proxy for the electric field amplitude in a region.}
\label{meas_tms_emf}
\end{figure}
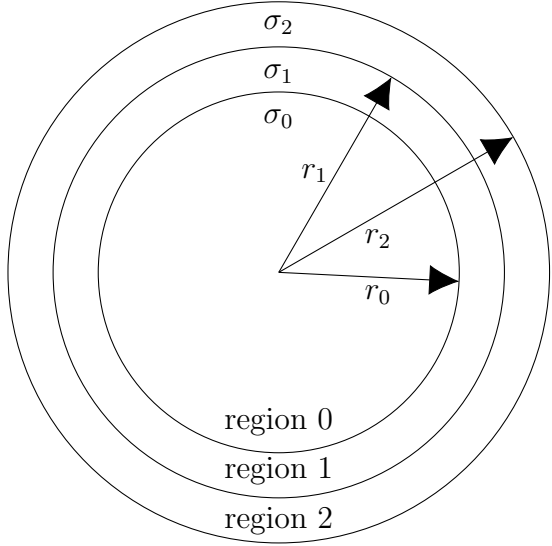

The electric fields of TEP and TMP, from which all results herein will be obtained, were derived by solving the quasistatic Maxwell Equations in terms of a vector spherical harmonic representation.  Figure \ref{meas_tms_emf} depicts the three-shell spherical head model which will be used in the derivation 
of the TEP and TMP electric fields. The spherical head of volume $V$ consists of three conducting spherical 
shells in which the regions from outermost to innermost are the scalp (region 2), skull (region 1) and brain 
(region 0) respectively with scalar conductivities $\sigma_i$ ($i = 0, 1, 2$). Reasonable estimates for the radii 
of the three shell model corresponding to human anatomy are $r_0 =$ 80 mm, $r_1 =$ 86 mm and  $r_2 =$ 92 mm 
\cite{Guilherme}. Typical values of the conductivities which will be used here are such that $\sigma_0 = \sigma_2$ 
and $\sigma_1/\sigma_0 = 1/80$ \cite{Rush_1} although, for sake of generality, these values will not be enforced 
initially.

In TMP the electric field arises from a current density ${\bf J}({\bf x},t)$ within a coil, supported external to $V$ only, driven 
by a current source. In TEP the electric field arises from an electric current density ${\bf J}({\bf x},t)$ in electrical contact 
with the external boundary of the scalp region. Regardless of the method the electric field is given at all 
positions ${\bf x}$ and times $t$ by \cite{JDJ}:
\begin{equation}
  {\bf E}({\bf x},t) = - \nabla \Phi({\bf x},t) - \frac{1}{c} \frac{\partial{\bf A}({\bf x},t)}{\partial t}
\label{eq_2.4a}
\end{equation}
where $ \Phi$ is the scalar potential and ${\bf A}$ is the vector potential. It will be convenient in this work 
to use the nondimensional position vector ${\bf r} =  {\bf x}/r_2$. The electric field is then given everywhere by:
\begin{equation}
  {\bf e}({\bf r},t) = - \frac{1}{r_2} \nabla \phi({\bf r},t) - \frac{1}{c} \frac{\partial{\bf a}({\bf r},t)}{\partial t}
\label{eq_2.4}
\end{equation}
where the derivatives of the $\nabla$ operator are with respect to the components of ${\bf r}$ and where the field
quantities are given by
${\bf e}({\bf r},t) = {\bf E}({\bf r}r_2,t)$, $\phi({\bf r},t) = \Phi({\bf r}r_2,t)$, ${\bf a}({\bf r},t) = 
{\bf A}({\bf r}r_2,t)$ and ${\bf j}({\bf r},t) = {\bf J}({\bf r}r_2,t)$.  

Given the frequencies of interest (less than 100kHz) we will make the usual quasistatic approximations. Under 
these approximations: (1) The scalar potential $\phi$ within $V$ obeys Laplace's Equation $\nabla^2 \phi = 0$, 
(2) Polarization and magnetization currents can be ignored so that the current within $V$ is Ohmic only 
(${\bf j}({\bf r},t) = \sigma({\bf r}) {\bf e}({\bf r},t)$ where $\sigma({\bf r})$ is the conductivity), (3) The 
vector potential within $V$ depends only upon currents external to $V$ since the secondary Ohmic currents, 
established within $V$ due to the electric field caused by the time varying external current, are relatively 
small by comparison and (4) The boundary conditions at the interface between regions $n$ and $n+1$ are 
$\hat{{\bf r}} \cdot {\bf j}_n = \hat{{\bf r}} \cdot {\bf j}_{n+1}$ and $\hat{{\bf r}} \times {\bf e}_n = \hat{{\bf r}} 
\times {\bf e}_{n+1}$ where $\hat{{\bf r}}$ is unit vector in the radial direction of a spherical coordinate system. 
The first boundary condition is a consequence of the quasistatic condition $\nabla \cdot {\bf j} = 0$ whereas the 
second boundary condition is valid in general.

\subsection{TEP}
To calculate ${\bf e}({\bf r},t)$ within $V$ we require a convenient form of the scalar potential 
$\phi({\bf r},t)$ and vector potential ${\bf a}({\bf r},t)$ suitable to the assumed spherical geometry. 
Given the geometry of the model a natural choice for representing the electric fields of both methods 
is the complete set of vector spherical harmonics. The electric scalar potential, obeying Laplace's 
Equation $\nabla^2 \phi = 0$ within $V$, can be written as sums of scalar spherical harmonics $Y_{jm}(\theta,\phi)$ in the three regions of the three-shell model as: 
\begin{eqnarray}
 \phi_0(r,\theta,\phi,t) &=&  r_2 \sum_{jm} A_{jm}(t) r^j Y_{jm}(\theta,\phi)
 \label{eq_1.0}
 \\
 \phi_1(r,\theta,\phi,t) &=&  r_2 \sum_{jm} B_{jm}(t) r^j Y_{jm}(\theta,\phi) + C_{jm}(t) r^{-(j+1)} Y_{jm}(\theta,\phi) 
 \label{eq_1.1}
 \\
 \phi_2(r,\theta,\phi,t) &=&  r_2 \sum_{jm} D_{jm}(t) r^j Y_{jm}(\theta,\phi) + E_{jm}(t) r^{-(j+1)} Y_{jm}(\theta,\phi) 
\label{eq_1.1b}
\end{eqnarray}
where the subscript $k=0,1,2$ of $\phi_k(r,\theta,\phi,t)$ denotes the region and where the indices of the double 
summation have values $j=0, \ldots, \infty$ and $m=-j, \ldots, j$. The quantities $A_{jm}$, $B_{jm}$, $C_{jm}$, 
$D_{jm}$ and $E_{jm}$ will be determined by the boundary conditions. Since the vector potential can be neglected
in the TEP case (the vector potential due to current in $V$ is negligible) the electric field is given by 
${\bf e}(r,\theta,\phi,t) = - \frac{1}{r_2} \nabla \phi(r,\theta,\phi,t)$ and in the three regions:
\begin{eqnarray}
  {\bf e}_0(r,\theta,\phi) 
  &=&  - \sum_{jm} A_{jm} [j(2j+1)]^{1/2} r^{j-1} {\bf Y}^{j-1}_{jm}(\theta,\phi)  
  \label{eq_1.3}
  \\
  {\bf e}_1(r,\theta,\phi) 
  &=& - \sum_{jm} B_{jm}  [j(2j+1)]^{1/2} r^{j-1} {\bf Y}^{j-1}_{jm}(\theta,\phi)   
  \nonumber \\	
  &-& \sum_{jm} C_{jm} [(j+1)(2j+1)]^{1/2} r^{-(j+2)} {\bf Y}^{j+1}_{jm}(\theta,\phi)      
  \label{eq_1.4}
  \\  
  {\bf e}_2(r,\theta,\phi) 
  &=& - \sum_{jm} D_{jm}  [j(2j+1)]^{1/2} r^{j-1} {\bf Y}^{j-1}_{jm}(\theta,\phi)
  \nonumber \\
  &-& \sum_{jm} E_{jm} [(j+1)(2j+1)]^{1/2} r^{-(j+2)} {\bf Y}^{j+1}_{jm}(\theta,\phi)      
\label{eq_1.4b}
\end{eqnarray}
where the time dependence has been suppressed for the sake of a compact notation.

The boundary conditions at $r=1$, $r=\alpha_1 = r_1/r_2$ and $r= \alpha_0 = r_0/r_2$ are 
\begin{eqnarray}
  \sigma_2 {\bf e}_2(1,\theta,\phi) \cdot {\hat{\bf r}} = {\bf j}(1,\theta,\phi) \cdot {\hat{\bf r}}
  \qquad \qquad \qquad \qquad \:\:
  \label{eq_1.5} 
  \\
  \sigma_1 {\bf e}_1(\alpha_1,\theta,\phi) \cdot {\hat{\bf r}} 
  = \sigma_2 {\bf e}_2(\alpha_1,\theta,\phi) \cdot {\hat{\bf r}}
  \qquad 
  {\hat{\bf r}} \times {\bf e}_1(\alpha_1,\theta,\phi)
  = {\hat{\bf r}} \times {\bf e}_2(\alpha_1,\theta,\phi) 
  \label{eq_1.6} 
  \\
  \sigma_0 {\bf e}_0(\alpha_0,\theta,\phi) \cdot {\hat{\bf r}} 
  = \sigma_1 {\bf e}_1(\alpha_0,\theta,\phi) \cdot {\hat{\bf r}}
  \qquad 
  {\hat{\bf r}} \times {\bf e}_0(\alpha_0,\theta,\phi)
  = {\hat{\bf r}} \times {\bf e}_1(\alpha_0,\theta,\phi)
  \label{eq_1.6b}
\end{eqnarray}

By applying these five boundary conditions we obtain a system of five linear equations which can be 
solved (see appendix \ref{solv_tes}) for the quantities $A_{jm}$, $B_{jm}$, $C_{jm}$, $D_{jm}$ and 
$E_{jm}$. Defining $\epsilon = \sigma_1/\sigma_0$ and making the reasonable assumption that 
$\sigma_2 = \sigma_0$ the following solution is obtained:
\begin{eqnarray}
  A_{jm} 
  &=& a_j \alpha_0^{-(2j+1)} \alpha_1^{-(2j+1)} {\mathcal D}_j^{-1} I_{jm}
  \nonumber \\
  B_{jm} 
  &=& b_j \alpha_0^{-(2j+1)} \alpha_1^{-(2j+1)} {\mathcal D}_j^{-1} I_{jm}
  \nonumber \\
  C_{jm} 
  &=& c_j \alpha_1^{-(2j+1)} {\mathcal D}_j^{-1} I_{jm}
  \nonumber \\
  D_{jm} 
  &=& [d_{0j} \alpha_0^{-(2j+1)} + d_{1j} \alpha_1^{-(2j+1)}] \alpha_1^{-(2j+1)} {\mathcal D}_j^{-1} I_{jm}  
  \nonumber \\
  E_{jm} 
  &=& e_j [\alpha_0^{-(2j+1)} - \alpha_1^{-(2j+1)}] {\mathcal D}_j^{-1} I_{jm}
  \label{eq_1.20i}
\end{eqnarray} 
where, for the sake of compact notation, we have defined 
\begin{eqnarray}
  a_j &=& \epsilon (2j+1)^2 
  \nonumber \\
  b_j &=& (2j+1)([1 + \epsilon]j + \epsilon)] 
  \nonumber \\
  c_j &=& -(1 - \epsilon)j(2j+1)
  \nonumber \\
  d_{0j} &=& ([1+ \epsilon]j+ \epsilon)([1+ \epsilon]j+ 1)   
  \nonumber \\
  d_{1j} &=& - (1 - \epsilon)^2 j(j+1)   
  \nonumber \\
  e_j &=& (1-\epsilon) j ([1 + \epsilon]j + \epsilon)]  
  \label{eq_1.20j}
\end{eqnarray} 
and
\begin{eqnarray}
  {\mathcal D}_j  
  &=& 
    - \epsilon j^3 \alpha_0^{-(2j+1)} \alpha_1^{-(2j+1)}  
    - \epsilon^2 j^2 (j+1) \alpha_0^{-(2j+1)} \alpha_1^{-(2j+1)}   
    - \epsilon j^2 (j+1) \alpha_1^{-(4j+2)}  
    \nonumber \\
    &+& \epsilon^2 j^2 (j+1) \alpha_1^{-(4j+2)}  	
    - j^2 (j+1) \alpha_0^{-(2j+1)} \alpha_1^{-(2j+1)} 
    - \epsilon j (j+1)^2 \alpha_0^{-(2j+1)} \alpha_1^{-(2j+1)}  
    \nonumber \\
    &+& j^2 (j+1) \alpha_1^{-(4j+2)}  
    - \epsilon j^2 (j+1) \alpha_1^{-(4j+2)}  
    - \epsilon j^2 (j+1) \alpha_0^{-(2j+1)}   
    \nonumber \\
    &-&  \epsilon^2 j(j+1)^2 \alpha_0^{-(2j+1)}  
    - \epsilon j(j+1)^2 \alpha_1^{-(2j+1)} 
    + \epsilon^2 j(j+1)^2 \alpha_1^{-(2j+1)}   
    \nonumber \\
    &+& j^2 (j+1) \alpha_0^{-(2j+1)}  
    +  \epsilon j(j+1)^2 \alpha_0^{-(2j+1)} 
    - j^2 (j+1) \alpha_1^{-(2j+1)}  
    \nonumber \\ 
    &+& \epsilon j^2 (j+1) \alpha_1^{-(2j+1)} 
  \label{eq_1.18b3}
\end{eqnarray}
and 
\begin{equation}
  I_{jm} = \frac{1}{\sigma_2} \int_0^{2\pi} \int_0^{\pi} {\bf j}(1,\theta,\phi) 
  \cdot {\hat{\bf r}} \; Y^*_{jm}(\theta,\phi) \sin \theta d\theta d\phi.
\label{eq_1.9}
\end{equation}
Note that since $\nabla \cdot {\bf j} = 0$ for a quasistatic system then, according to Gauss's Law, 
$I_{00} = 0$ therefore the indices of the double summation are now $j=1, \ldots, \infty$ and $m=-j, \ldots, j$. 

Considering the solutions for the electric field within the brain as given by equation (\ref{eq_1.3}) together 
with equation (\ref{eq_1.9}) it is apparent that the electric field is independent of the size of the head 
$r_2$. Therefore for two different size three-shell model heads, with the same relative size shells, the 
electric fields will be identical at any given nondimensional position within the head if the current densities
${\bf j}(1,\theta,\phi)$ are identical. However for a fixed angular distribution of current density, since the
surface area of the electrodes increases as $r_2^2$, then so does the total current delivered to the electrodes 
by the current source. 

\subsection{TMP}

In accordance with equation (\ref{eq_2.4}) both the scalar potential $\phi({\bf r},t)$ and vector potential 
${\bf a}({\bf r},t)$ must be considered to obtain the TMP electric field ${\bf e}({\bf r},t)$ within $V$. 
In terms of the dimensionless spatial coordinate $\bf r$ the vector potential in the quasistatic case is given by 
\cite{JDJ}
\begin{equation}
  {\bf a}({\bf r},t) = \frac{r_2^2}{c} \iiint \frac{{\bf j}({\bf r}',t)}{|{\bf r} - {\bf r}'|} d^3{\bf r}'. 
  \label{vec_pot}
\end{equation}
Expanding the integrand in terms of vector spherical harmonics (see \cite{VMK} pg 229) we can write
the electric field within $V$ as  
\begin{equation}
  {\bf a}({\bf r},t) = c \sum\limits_{ljm} \frac{r^l}{2l+1} {\bf Y}^l_{jm}(\theta,\phi)  J^l_{jm}(t) 
\end{equation}
where
\begin{equation}
  J^l_{jm}(t) = \frac{4\pi r_2^2}{c^2} \int \int \int \frac{1}{r'^{l+1}}
  {\bf j}(r',\theta',\phi',t) \cdot {\bf Y}^{*l}_{jm}(\theta',\phi')  r'^2 \sin \theta' dr' d\theta' d\phi'.
\label{eq_2.3}
\end{equation}

Since the ohmic current within $V$ can be neglected in the calculation of the vector potential 
then it follows that $\nabla \times {\bf b} = \nabla \times \nabla \times {\bf a} = 0$ within
$V$. Applying this constraint (making use of identities given in \cite{VMK} pg 217) one finds that 
$J^{j+1}_{jm} = 0$ and therefore
\begin{equation}
  {\bf a}({\bf r},t) = c \sum_{jm} \frac{r^j}{2j+1} J^j_{jm}(t) {\bf Y}^j_{jm}(\theta,\phi)
  + c \sum_{jm} \frac{r^{j-1}}{2j-1} J^{j-1}_{jm}(t) {\bf Y}^{j-1}_{jm}(\theta,\phi) 
\label{eq_2.2}
\end{equation}

Since the quasistatic vector potential given by equation (\ref{vec_pot}) satisfies $\nabla \cdot {\bf a} =  0$ 
everywhere and since $\nabla \cdot {\bf e} = 0$ within $V$ then according to equation (\ref{eq_2.4}) the scalar 
potential must satisfy the Laplace equation within $V$. Therefore within $V$ the scalar potential in the three 
regions can written as:
\begin{eqnarray}
  \phi_0(r,\theta,\phi) &=& r_2 \sum_{jm} A_{jm} r^j Y_{jm}(\theta,\phi)
  \label{eq_2.5}
  \\
  \phi_1(r,\theta,\phi) &=& r_2 \sum_{jm} B_{jm} r^j Y_{jm}(\theta,\phi) 
  + C_{jm} r^{-(j+1)} Y_{jm}(\theta,\phi) 
  \label{eq_2.6}
  \\
  \phi_2(r,\theta,\phi) &=& r_2 \sum_{jm} D_{jm} r^j Y_{jm}(\theta,\phi) 
  + E_{jm} r^{-(j+1)} Y_{jm}(\theta,\phi) 
  \label{eq_2.6a}
\end{eqnarray}
and the corresponding electric fields are:
\begin{eqnarray}
  {\bf e}_0(r,\theta,\phi) &=& - \sum_{jm} [j(2j+1)]^{1/2} A_{jm} r^{j-1} {\bf Y}^{j-1}_{jm}(\theta,\phi)
  \nonumber \\
  &-& \sum_{jm} \frac{r^j}{2j+1} \frac{\partial J^j_{jm}}{\partial t} {\bf Y}^j_{jm}(\theta,\phi)
  - \sum_{jm} \frac{r^{j-1}}{2j-1} \frac{\partial J^{j-1}_{jm}}{\partial t} {\bf Y}^{j-1}_{jm}(\theta,\phi) 	
  \label{eq_2.6e}
  \\
  {\bf e}_1(r,\theta,\phi) &=&  
  - \sum_{jm} [j(2j+1)]^{1/2} B_{jm} r^{j-1} {\bf Y}^{j-1}_{jm}(\theta,\phi)
  \nonumber \\	
  &-& \sum_{jm} [(j+1)(2j+1)]^{1/2} C_{jm} r^{-(j+2)} {\bf Y}^{j+1}_{jm}(\theta,\phi) 
  \nonumber \\
  &-& \sum_{jm} \frac{r^j}{2j+1} \frac{\partial J^j_{jm}}{\partial t} {\bf Y}^j_{jm}(\theta,\phi)
  - \sum_{jm} \frac{r^{j-1}}{2j-1} \frac{\partial J^{j-1}_{jm}}{\partial t} {\bf Y}^{j-1}_{jm}(\theta,\phi)
  \label{eq_2.6f}
  \\
  {\bf e}_2(r,\theta,\phi) &=&  
  - \sum_{jm} [j(2j+1)]^{1/2} D_{jm} r^{j-1} {\bf Y}^{j-1}_{jm}(\theta,\phi)
  \nonumber \\	
  &-& \sum_{jm} [(j+1)(2j+1)]^{1/2} E_{jm} r^{-(j+2)} {\bf Y}^{j+1}_{jm}(\theta,\phi) 
  \nonumber \\
  &-& \sum_{jm} \frac{r^j}{2j+1} \frac{\partial J^j_{jm}}{\partial t} {\bf Y}^j_{jm}(\theta,\phi)
  - \sum_{jm} \frac{r^{j-1}}{2j-1} \frac{\partial J^{j-1}_{jm}}{\partial t} {\bf Y}^{j-1}_{jm}(\theta,\phi)
  \label{eq_2.6g}
\end{eqnarray}
Note that once again the time dependence of the fields has been suppressed for the sake of compact notation.

The boundary conditions are still those given by equations (\ref{eq_1.5})-(\ref{eq_1.6b}) albeit with 
${\bf j} \cdot {\bf n} = 0$ and now the electric field has a component due to induction as well as that 
due to surface charges. By applying the five boundary conditions we again obtain a system of five linear 
equations which can be solved (see appendix \ref{solv_tms}) for the quantities $A_{jm}$, $B_{jm}$, 
$C_{jm}$, $D_{jm}$ and $E_{jm}$. For all three regions the following simple solution is obtained:
\begin{eqnarray}
  {\bf e}(r,\theta,\phi) = - \sum_{jm} \frac{r^j}{2j+1} \frac{\partial J^j_{jm}}{\partial t} 
  {\bf Y}^j_{jm}(\theta,\phi)	
  \label{eq_2.85}
\end{eqnarray}
Note that the electric field due to the surface charge exactly cancels the $l=j-1$ components of the 
magnetically induced components of the electric field and therefore the conductivities do not appear 
anywhere in the solution. Nondimensional spatial coordinates have been used in equation (\ref{eq_2.85}) and 
the only quantity which depends on the size of the head $r_2$ is $J^j_{jm}$ as defined in equation 
(\ref{eq_2.3}). Accordingly as $r_2$ is decreased the current density must increase as $r_2^2$ in order 
to achieve the same electric field magnitude at the nondimensional radial position $r$ within the head. 
This presents a challenge for creating small animal TMP systems with electric fields of angular resolution 
and magnitude comparable to those in humans. If smaller coils are used to try to achieve angular resolution
comparable to that in humans the resistance of such coils will, for frequencies of interest here, increase 
approximately as $r_2^{-2}$ while the current needed to obtain similar electric fields in the cortex is 
unchanged. As a result the power dissipated in the coil will increase approximately as $r_2^{-1}$ demanding
efficient and relatively small cooling systems to prevent damage to the TMP coil. Of course a more complete
description of the differences between humans and small animals would include differences in the size of 
each shell of the three shell model and the conductivities therein.

One important point to note is that {\it TEP and TMP electric fields within the brain region exist in 
orthogonal subspaces}. This follows from the VSH property 
$\int_0^{\pi} \int_0^{\pi} {\bf Y}^l_{jm}(\theta,\phi) \cdot {\bf Y}^{l'}_{j'm'}(\theta,\phi) 
\sin \theta d\theta d\phi = \delta_{ll'} \delta_{jj'} \delta_{mm'}$ and from equations  (\ref{eq_1.3}) and (\ref{eq_2.6e})  which show that the TEP and TMP brain electric fields are spanned by the $l=j-1$ and $l=j$ 
VSH components respectively. Also note that the TMP electric field, unlike the TEP field, has no radial 
component (see appendix \ref{VSH_props}). The orthogonality of the TEP and TMP fields has great consequence 
since the coupling of the electric field to neurons is dependent upon the relative direction of the field 
and the neuronal fibers. Consequently even if the electric field of TEP and TMP are angularly "focused" on 
the same regions of the cortex completely different populations of neurons may be affected by each. This 
may be of particular importance to studies which use suprathrsehold TMS to probe changes in cortical 
excitability due to TES.

\section{Results}

\label{power}

Here, estimates are given, in the context of the three-shell TEP/TMP model, for select metrics of the relative power dissipated in the scalp and brain regions. In addition a simple estimate of the relative power utilized by the methods is presented to give a balanced understanding of the limitations and strengths of each.

Three power metrics are calculated: $R^{tep}$, $R^{tmp}$ and $R$. The quantities 
$R^{tep}$ and $R^{tmp}$ are the scalp-to-brain ratios of power dissipation for the TEP and TMP cases respectively. 
These quantities enable one to estimate the power dissipated in the scalp for a given power dissipated in the
brain. However the radial dependence of the TEP and TMP electric fields are fundamentally different making it
difficult to directly compare the relative energy dissipated (or mean-squared electric field) in the scalp for 
the two methods. To yield a better direct comparison the quantity $R$ is calculated which gives the TEP-to-TMP
ratio of scalp energy dissipation for the case of similar TEP and TMP electric fields at the radial position of 
the cortex. Example calculations of each ratio are given for the case of typical electrode and coil geometries. 
In addition the three quantities are calculated in the case where only one VSH of index $j$ contributes to the
field. This leads to the calculation of an asymptotic limit to the ratio $R$. Note that since the conductivities 
of the brain and scalp are taken to be equal in this three-shell model then the dissipated power ratios are
equivalent to mean-squared electric field ratios.

\subsection{Relative Power Dissipation in Brain and Scalp: TEP Case}
\label{tes_power}

In this subsection we calculate $R^{tep}$, the ratio of the spatiotemporal averaged power dissipated in the 
scalp and brain regions for the TEP electric field. The averaged power dissipated in region $k$ is given by:
\begin{eqnarray}  
  P_k 
  = \frac{\sigma_k}{V_k T} \int_0^T \int_{R_k} |{\bf e}(r,\theta, \phi,t)|^2 r^2 \sin \theta dr d\theta d\phi dt
  \label{eq_1.22x}
\end{eqnarray}
where $V_k$ is the volume of region $R_k$, $\sigma_k$ is its conductivity and $T$ is the temporal averaging interval. The interval $T$ could be any meaningful time interval for the temporal waveform of the current source.
For example, it could a period of a periodic waveform or it could be an interval which is large compared to such 
a period. Note that if the current density is separable with respect to the spatial and temporal variables (that is {\bf j}({\bf r},t) = I(t) {\bf f}({\bf r})) then temporal averaging is inconsequential since
the time dependence cancels in the ratio $R^{tep}$. All but one of the current densities considered in this work will be separable. The exceptions, as discussed in section \ref{nonsep}, will be a TMP system comprised of two circular coils and a TEP system comprised of two electrode pairs each driven by independent sinusoidal current sources of different frequencies. Also note that $P_k$ can also be interpreted as the product of the mean-square electric field and the conductivity for region $k$.

Using equations (\ref{eq_1.3}), (\ref{eq_1.20i}) and (\ref{eq_1.22x}) we can write the average power dissipated 
in region $0$ due to the TEP electric field as: 
\begin{eqnarray}  
  P^{tep}_0 
  &=& \frac{\sigma_0}{r_2^2 V_0} \sum_{jm} j \overline{|A_{jm}|^2} \alpha_0^{2j+1}   
  \nonumber \\
  &=& \frac{3\sigma_0}{4\pi r_2^2 \alpha_0^3} \sum_{jm} j a_j^2 \alpha_0^{-(2j+1)} \alpha_1^{-(4j+2)} 
  {\mathcal D}_j^{-2} \overline{|I_{jm}|^2} 
  \label{eq_1.22}
\end{eqnarray}
where the line over time dependent quantities denotes a time average.
Similarly the average power dissipated in region 2 is:
\begin{eqnarray}  
  P^{tep}_2 
  &=&  \frac{\sigma_0}{V_2}
  \sum_{jm} 
  \left[ 
    \overline{|D_{jm}|^2} j (r_2^{2j+1} - r_1^{2j+1})
    + \overline{|E_{jm}|^2} (j+1) (r_1^{-(2j+1)} - r_2^{-(2j+1)})
  \right]
  \nonumber \\ 
  &=& \frac{3\sigma_0}{4\pi (1-\alpha_1^3) r_2^2} \sum_{jm} 
  j [d_{0j} \alpha_0^{-(2j+1)} \alpha_1^{-(2j+1)} + d_{1j} \alpha_1^{-(4j+2)}]^2  {\mathcal D}_j^{-2} \overline{|I_{jm}|^2} 
  (1 - \alpha_1^{2j+1}) 
  \nonumber \\
  &+& \frac{3\sigma_0}{4\pi (1-\alpha_1^3)r_2^2} \sum_{jm} 
  (j+1) e_j^2 [\alpha_0^{-(2j+1)} - \alpha_1^{-(2j+1)}]^2 {\mathcal D}_j^{-2} \overline{|I_{jm}|^2} [\alpha_1^{-(2j+1)} - 1] 
  \label{eq_1.24}
\end{eqnarray}
Each term in the summation of equation (\ref{eq_1.22}) or equation (\ref{eq_1.24}) is the average power 
$P^{tep,k}_{jm}$ dissipated in the VSH component of the TEP electric field indexed by $(j,m)$ in regions 
$k=0,2$. The ratio, $R^{tep}= P^{tep}_2/P^{tep}_0$, of the average power dissipated in the scalp to that 
dissipated in the brain is then
\begin{eqnarray}  
  R^{tep}
  &=& \frac{\alpha_0^3}{1-\alpha_1^3} 
  \frac
  {\sum_{jm} j [d_{0j} \alpha_0^{-(2j+1)} \alpha_1^{-(2j+1)} + d_{1j} \alpha_1^{-(4j+2)}]^2  
  {\mathcal D}_j^{-2} \overline{|I_{jm}|^2} (1 - \alpha_1^{2j+1})}
  {\sum_{jm} j a_j^2 \alpha_0^{-(2j+1)} \alpha_1^{-(4j+2)} {\mathcal D}_j^{-2} \overline{|I_{jm}|^2}} 
  \nonumber \\
  &+& \frac{\alpha_0^3}{1-\alpha_1^3} 
  \frac
  {\sum_{jm} (j+1) e_j^2 [\alpha_0^{-(2j+1)} - \alpha_1^{-(2j+1)}]^2 {\mathcal D}_j^{-2} 
  \overline{|I_{jm}|^2} [\alpha_1^{-(2j+1)} - 1]}
  {\sum_{jm} j a_j^2 \alpha_0^{-(2j+1)} \alpha_1^{-(4j+2)} {\mathcal D}_j^{-2} \overline{|I_{jm}|^2}} 
  \label{eq_1.25a}
\end{eqnarray}

\begin{figure}[h]
    \input{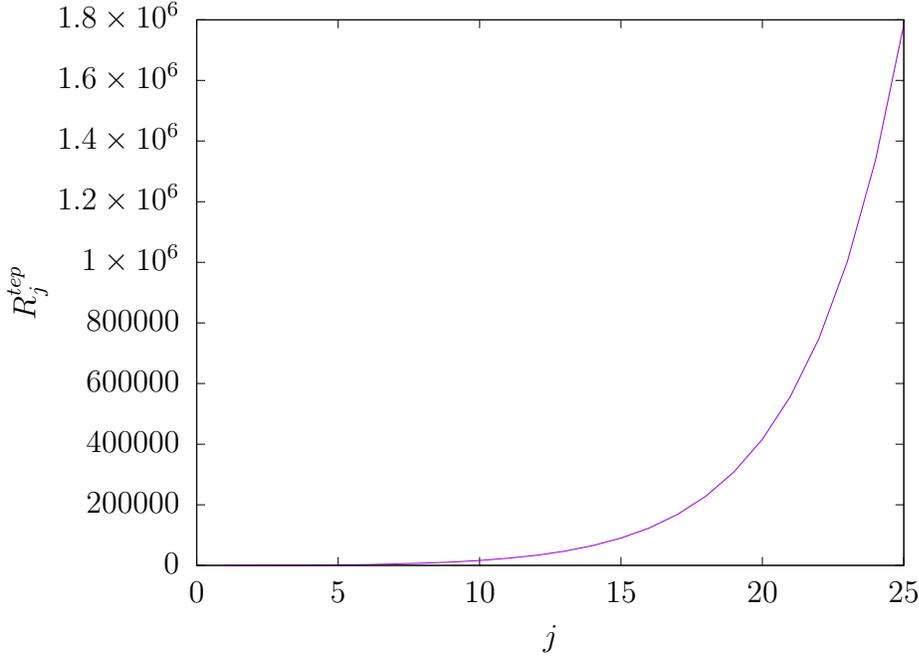}
    \caption{$R_j^{tep}$ versus $j$ for TEP electric fields. $R_j^{tep}$ is the ratio of the mean-squared electric field over the scalp region to that over the brain region for an electric field comprised of a single vector spherical harmonic component indexed by the pair of integers $j$ and $m$. The index $j$ of the vector spherical representation of the electric field is reciprocally related to spatial scale in the field. Note that $R_j^{tep}$ is independent of index $m$ for any given value of $j$. The cost of greater electric field focality in the brain is greater mean-squared electric field in the scalp relative to the brain.}
    \label{R_j_tep}
\end{figure}

In general $R^{tep}$ depends upon $I_{jm}$, that is, it depends upon the geometry of the TEP electrodes and the 
magnitude of the current supplied to the electrodes. However general features can can be elucidated by considering 
the separable case when the current source is such that $I_{jm} = 0$ for all but one value of $j$ ($m$ not restricted). In 
that case the power ratio for the $j^{th}$ component, $R^{tep}_j= P^{tep,2}_{jm}/P^{tep,0}_{jm}$, is given by:
\begin{eqnarray}  
  R^{tep}_j 
  &=& \frac{\alpha_0^3}{1 - \alpha_1^3}
  [(d_{0j}/a_j) \alpha_0^{-(2j+1)} + (d_{1j}/a_j) \alpha_1^{-(2j+1)}]^2  [1 - \alpha_1^{2j+1}] \alpha_0^{2j+1}
  \nonumber \\
  &+& \frac{\alpha_0^3}{1 - \alpha_1^3}
    \frac{j+1}{j} (e_j/a_j)^2 [\alpha_0^{-(2j+1)} - \alpha_1^{-(2j+1)}]^2 
    [1 - \alpha_1^{2j+1}] \alpha_0^{2j+1} \alpha_1^{2j+1}	
  \label{eq_1.26b}
\end{eqnarray}
Figure \ref{R_j_tep} shows the dependence of $R_j^{tep}$ upon $j$ for a three shell conductor model. It is 
clear from the figure that $R_j^{tep}$ increases as $j$ increases. In other words, \textit{as the spatial detail of 
the electric field increases (eg. more focality) so does the energy dissipated in the scalp relative 
to that dissipated in the brain}.

\begin{figure}[h]
\centering
\begin{tikzpicture}
  \coordinate (O) at (0,0);

  \shade[ball color = gray!10, opacity = 0.1] (0,0) circle [radius = 3cm];

  \draw (O) circle [radius=3cm];


  \draw[densely dashed] (O) to [edge label = $r_2$] (0,2.85);
  \draw[densely dashed] (O) -- (2.04,2.04);
  \draw (0.35,0.35) arc [start angle = 35, end angle = 85, x radius = 5mm, y radius = 5mm];
  \node at (.4,0.8) {$\beta$};

  \draw[densely dashed] (O) -- (0.8,2.85);
  \draw (0.3,1.1) arc [start angle = 35, end angle = 104, x radius = 2.7mm, y radius = 2.7mm];
  \node at (.2,1.6) {$\theta_o$};

  \draw[densely dashed] (2.04,2.04) -- (2.55, 1.50);
  \node at (2.7,1.95) {$r_e$};
  \draw[densely dashed] (0,2.86) -- (.75,2.86);
  \node at (0.45,3.3) {$r_e$};


  \draw (0,2.85) ellipse (23pt and 4pt);  
  \draw [rotate=-45] (0,2.85) ellipse (23pt and 4pt);


\end{tikzpicture}
\caption{Spherical head model with two TEP electrodes on the scalp surface ($r=r_2$). Each electrode (outlined) 
subtends the angle $\theta_o$ from its center and the centers of the two electrodes are separated by the angle $\beta$.}
\label{tep_electrodes}
\end{figure}
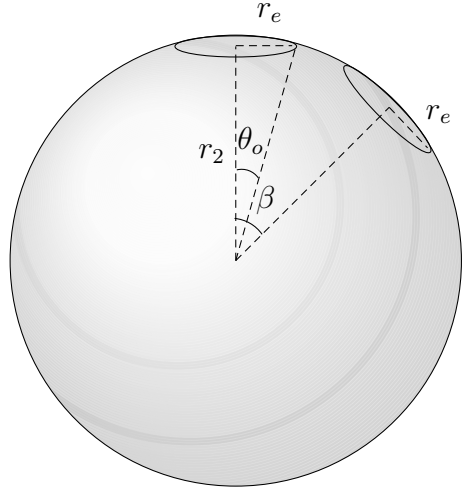

To estimate $R^{tep}$ for a typical TEP system consider the scalp electrode system depicted in figure 
\ref{tep_electrodes}. In this example the system is comprised of two electrodes each subtending an angle 
$\theta_o$ on the scalp surface, with electrode centers separated by the angle $\beta$. Appendix
\ref{I_coef} derives the $I_{jm}$ for such a system which is found to be:
\begin{eqnarray}
  I_{jm} &=& I^+_{j0} \left [ \delta_{m0} - \sqrt{\frac{(j-m)!}{(j+m)!}} P^m_j(\cos \beta) \right]
  \nonumber \\
  &=& I^+_{j0} \left [ \delta_{m0} - \sqrt{\frac{4\pi}{2j+1}} {\tilde P}^m_j(\cos \beta) \right]
  \label{eq_ex_9.10}
\end{eqnarray}
where
\begin{eqnarray}
  I^+_{j0} = 2 \pi I_o \sqrt{\frac{1}{2j+1}}
  \left[ 
    \sqrt{\frac{1}{2j+3}} \tilde{P}_{j+1}(\cos \theta_o) - \sqrt{\frac{1}{2j-1}} \tilde{P}_{j-1}(\cos \theta_o)
  \right]
  \label{eq_ex_9.11}
\end{eqnarray}
and where $I_o$ is the radial component of a uniform current density provided by the electrodes.

\begin{figure}[ht]
    \input{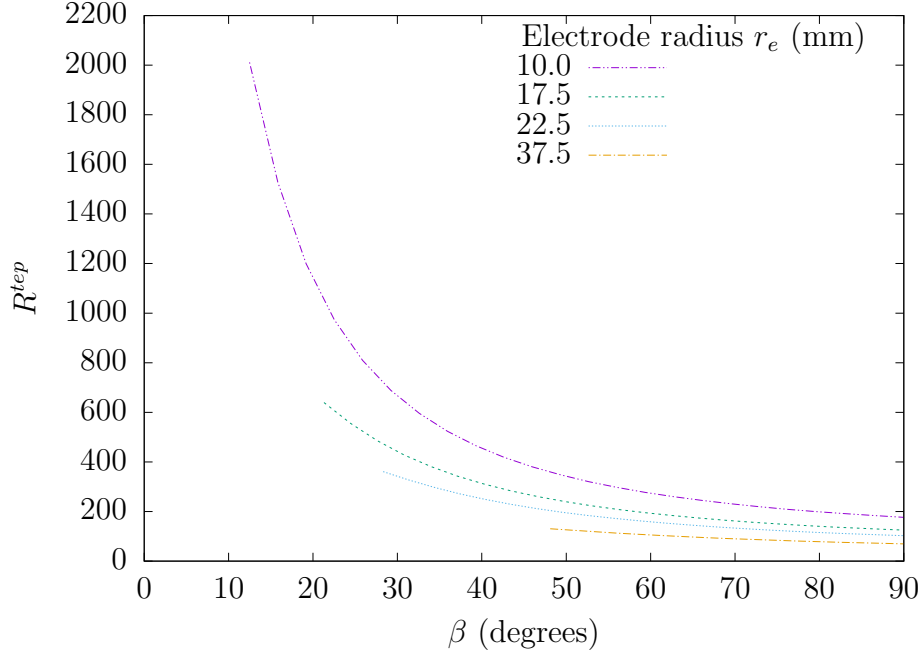}
    \caption{A plot of ratio $R^{tep}$ versus electrode separation angle $\beta$ for four different electrode 
    radii $r_e$. $R^{tep}$ is the ratio of the mean-squared electric field over the scalp region to that of the brain region for the specific electrode system depicted in figure \ref{tep_electrodes}. The angle $\theta_o$ which subtends the electrode from its center is determined according to
    $\theta_o = \cos^{-1} \sqrt{1-r_e^2}$. Each plot extends over the range $\beta = [\theta_o, 90]$ so that
    the electrodes do not overlap.}
    \label{R_tep}
\end{figure}

C++ computer code (available upon request) was written to perform all summations within this work. The computation
of the normalized associated Legendre functions $P^m_j$ was adopted from a standard reference \cite{NRC}. Figure 
\ref{R_tep} shows the dependence of $R^{tep}$ versus electrode separation angle $\beta$ for four different electrode 
sizes $\theta_o$. $R^{tep}$ increase as spatial detail increases with smaller electrodes or smaller separation between
the electrodes.

\subsection{Relative Power Dissipation in Brain and Scalp: TMP Case}
\label{tms_power}

\begin{figure}[ht]
    \input{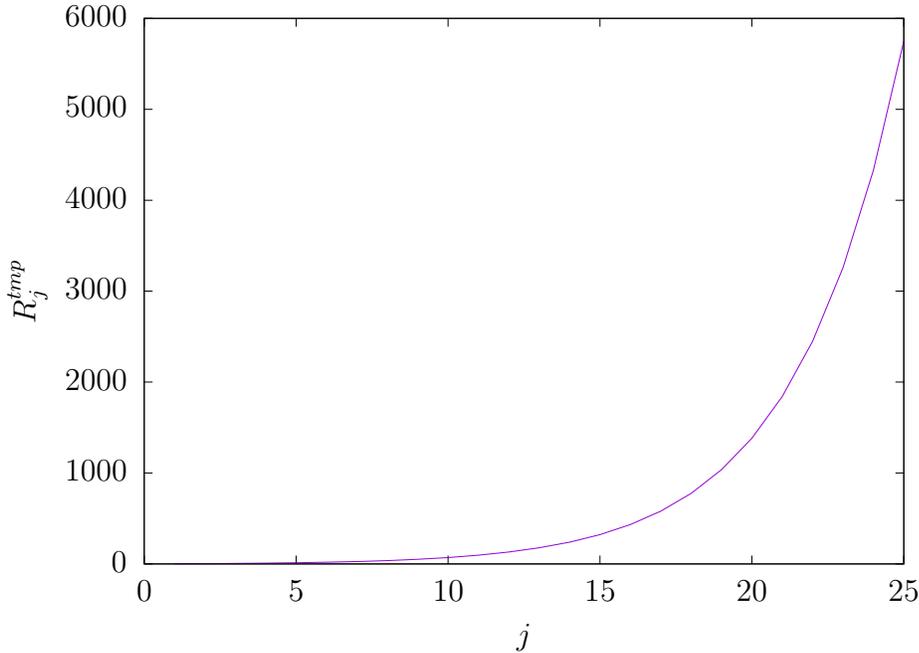}
    \caption{$R_j^{tmp}$ versus $j$ for TMP electric fields. $R_j^{tmp}$ is the ratio of the mean-squared electric field over the scalp region to that over the brain region for an electric field comprised of a single vector spherical harmonic component indexed by the pair of integers $j$ and $m$. The index $j$ of the vector spherical representation of the electric field is reciprocally related to spatial scale in the field. Note that $R_j^{tmp}$ is independent of index $m$ for any given value of $j$. As with TEP the cost of greater electric field focality in the brain is greater mean-squared electric field in the scalp relative to the brain. However that cost is much greater for TEP as compared to TMP.}
    \label{R_j_tmp}
\end{figure}

In this subsection we calculate the ratio of the spatially-averaged power dissipated in the brain and 
scalp regions for the TMP electric field. Using equations (\ref{eq_2.85}) and (\ref{eq_1.22x}) we can 
write the average power dissipated in region $0$ as: 
\begin{eqnarray}
  P^{tmp}_0 
  = \frac{3 \sigma_0}{4\pi \alpha_0^3} \sum_{jm}
  \frac{\alpha_0^{2j+3}}{(2j+1)^2 (2j+3)} 
  \overline{\left| \frac{\partial{J}^j_{jm}}{\partial t} \right|^2} 
  \label{eq_2.20b} 
\end{eqnarray}
Similarly the average dissipated power in region 2 is:
\begin{eqnarray}
  P^{tmp}_2 
  = \frac{3\sigma_0}{4\pi (1-\alpha_1^3)} \sum_{jm}
  \frac{1 - \alpha_1^{2j+3}}{(2j+1)^2 (2j+3)} 
  \overline{\left| \frac{\partial{J}^j_{jm}}{\partial t} \right|^2} 
  \label{eq_2.21} 
\end{eqnarray}
and the ratio $R^{tmp} = P_2^{tmp}/P_0^{tmp}$ is:
\begin{eqnarray}
  R^{tmp} =  
  \frac{\alpha_0^3}{1 - \alpha_1^3}
  \frac{\sum_{jm} (1 - \alpha_1^{2j+3})[(2j+1)^2 (2j+3)]^{-1} 
  \overline{\left| \frac{\partial{J}^j_{jm}}{\partial t} \right|^2}}
  {\sum_{jm} \alpha_0^{2j+3}[(2j+1)^2 (2j+3)]^{-1} 
  \overline{\left| \frac{\partial{J}^j_{jm}}{\partial t} \right|^2}}  
  \label{eq_2.22b} 
\end{eqnarray}
If the current source is such that $J_{jm} = 0$ for all but one value of $j$ ($m$ not restricted) then 
$R^{tmp}_j = P^{tmp,2}_{jm}/P^{tmp,0}_{jm}$ is
\begin{eqnarray}
  R^{tmp}_j =  
  \frac{\alpha_0^3}{1 - \alpha_1^3}
  \frac{1 - \alpha_1^{2j+3}}
  {\alpha_0^{2j+3} }  
  \label{eq_2.22d} 
\end{eqnarray}
Figure \ref{R_j_tmp} shows the dependence of $R_j^{tmp}$ upon $j$.
Again, as in the case of TEP, the spatial detail of the electric field comes at a cost. The energy dissipated 
in the scalp relative to the energy dissipated in the brain increases as the electric field is made more 
spatially detailed. However, in contrast to the TEP electric field, $R_j^{tmp}$ is much smaller than $R_j^{tep}$
for a given $j$. For example at $j=20$ $R_j^{tep}$ is approximately 35 times greater than $R_j^{tmp}$. 

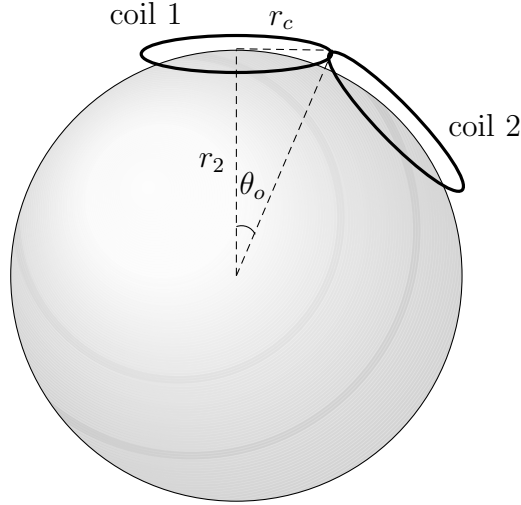
\begin{figure}
\centering
  \begin{tikzpicture}
    \coordinate (O) at (0,0);

    \shade[ball color = gray!10, opacity = 0.1] (0,0) circle [radius = 3cm];

    \draw (O) circle [radius=3cm];

    \draw[densely dashed] (O) to [edge label = $r_2$] (0,3.01);
    \draw[densely dashed] (O) -- (1.28,3.0);

    \draw (0.25,0.55) arc [start angle = 25, end angle = 110, x radius = 2mm, y radius = 2mm];
    \node at (.22,1.2) {$\theta_o$};

    \draw [very thick] (0,2.95) ellipse (36pt and 7pt); 
    \node at (-1.2,3.5) {coil 1};
    \draw [rotate=-46][very thick] (0,2.95) ellipse (36pt and 7pt); 
    \node at (3.3,2.0) {coil 2};

    \draw[densely dashed] (0,3.02) -- (1.28,3.0);
    \node at (.6,3.4) {$r_c$};

  \end{tikzpicture}
  \caption{Spherical head model with two circular TMP coils with radii $r_c$. The centers of coil 1 and 2 are on the scalp
  scalp surface ($r=r_2$).
  Note that the angle between the planes of the two coils is $\pi - 2 \theta_o$. }
  \label{tmp_coil}
\end{figure}

\begin{figure}[ht]
    \scalebox{1.2}{\input{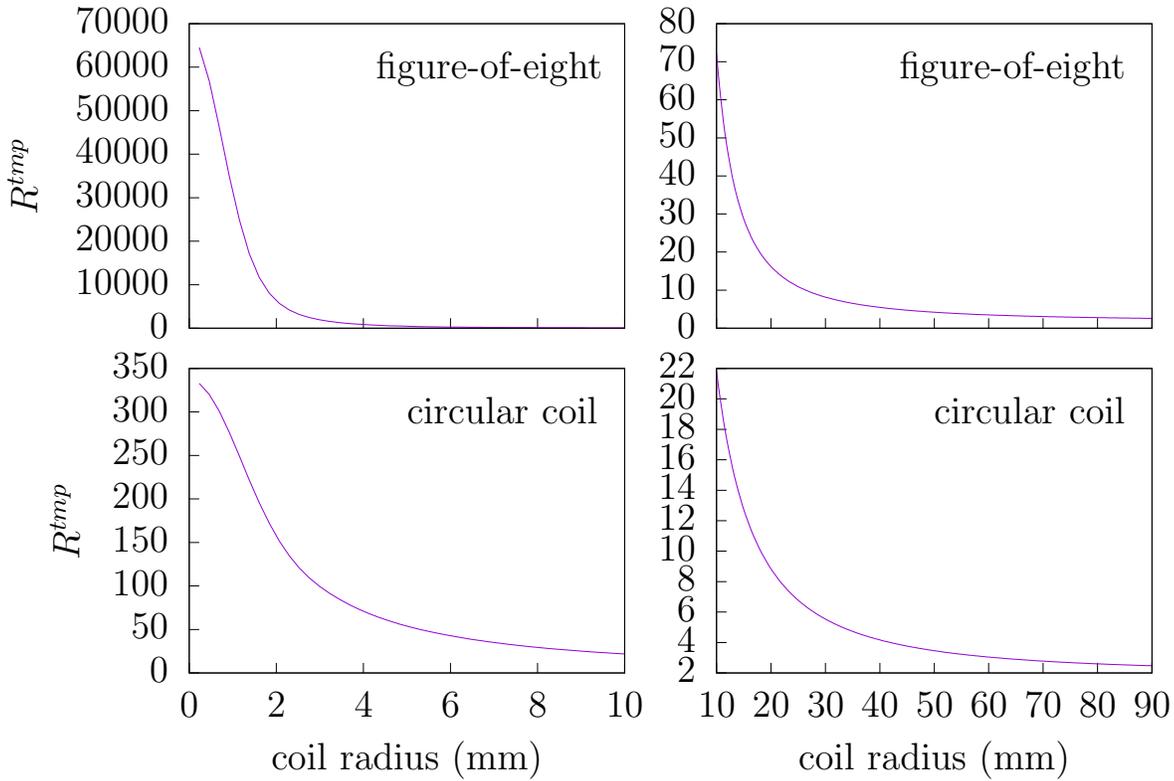}}
    \caption{The ratio $R^{tmp}$ versus coil radius $r_c$ for circular coils (bottom row) and 
    figure-8 (top row) coil. $R^{tmp}$ is the ratio of the mean-squared electric field over the scalp region to that of the brain region for the coil geometries depicted in figure \ref{tmp_coil}. For small radii the difference between the values of $R^{tmp}$ for the circular 
    and figure-8 coils is quite pronounced. Therefore, to draw attention to this difference, the left plots 
    cover 0-10 mm radii whereas the right plots cover 10-90 mm.}
    \label{R_tmp}
\end{figure}

The value of $R^{tmp}$ will of course depend on the geometry of the TMP coil. That is it will depend upon $J_{jm}$.
Here calculations of $R^{tmp}$ are given for simple thin circular TMS coils and figure-8 coils as depicted 
in figure \ref{tmp_coil}. The specifications for TMS coils, which typically contain many windings of Litz wire, are 
usually given in terms of an inner and outer radius for the winding. Here the coils are approximated by a single 
winding at the average of typical inner and outer radii. For the circular coil (coil 1 of figure \ref{tmp_coil}) 
assume the current density $\bf j$ is a thin ring of current of amplitude $I(t)$ and radius $r_c$ (in units of $r_2$) 
inscribed on a plane tangent to the outer surface of the scalp region and centered on the vertical axis. Appendix 
\ref{J_coef} calculates $J^j_{jm}$ for this simple coil to be: 
\begin{equation}
  J^{j}_{jm} 
  = i \delta_{m0}\frac{8\pi^2 r_2^2}{c^2 \rho_o^j} I \sqrt{1 - \cos^2 \theta_o} 
  {\tilde P}^{1}_j(\cos \theta_o)
  \label{eq_2.220}
\end{equation} 
where $\theta_o = \cos^{-1}(1/\sqrt{r_c^2 + 1})$. A figure-8 coil can be constructed from two circular coils (coils 1 and 2 of figure \ref{tmp_coil}) with currents circulating in opposite senses and with coil 2 rotated by an angle $2 \theta_o$ relative to coil 1. For this figure-8 coil $J^{j}_{jm} = J^{j+}_{jm} - J^{j-}_{jm} $ where $J^{j+}_{jm}$ and $J^{j-}_{jm}$ are contributions from coil 1 and 2 respectively. Appendix \ref{J_coef} calculates $J^j_{jm}$ for this figure-8 coil to be: 
\begin{equation}
  J^{j}_{jm} 
  = i \frac{8\pi^2 r_2^2}{c^2 \rho_o^j} I \sqrt{1 - \cos^2 \theta_o} 
  \left[
    \delta_{m0} - (-1)^m \sqrt{\frac{4\pi}{2j+1}} {\tilde P}_j^m(\cos 2\theta_o) 
  \right]
  {\tilde P}^{1}_j(\cos \theta_o)
  \label{eq_2.221}
\end{equation} 
Figure \ref{R_tmp} gives a plot of $R^{tmp}$ versus coil radii for the circular and figure-8 coils. Clearly the 
value of $R^{tmp}$ increases as the radius of the coil decreases. Note that for coil radii less than 10 mm the difference
between the figure-8 $R^{tmp}$ and circular coil $R^{tmp}$ is quite large. However for coil radii greater than
10 mm the difference is not near as stark.

The quantities $R^{tep}$ and $R^{tmp}$ are useful for estimating the mean-squared electric field in the scalp 
(brain) given an estimate for mean-squared electric field in the brain (scalp). Even though the plots of 
$R^{tep}$ and $R^{tmp}$ given in figures \ref{R_tep} and \ref{R_tmp} respectively show that the values of 
$R^{tmp}$ are typically much smaller than $R^{tep}$ for standard TEP electrode and TMP coil configurations a 
direct comparison of these quantities may be inadequate for estimating the relative intensities of the TEP 
and TMP scalp electric fields. This direct comparison is complicated by the fact that the electric fields of 
TEP and TMP have different radial dependences which may skew the volume averages over the brain region.
Furthermore, while it is the scalp electric field that often limits the brain electric field amplitude, the 
usual target of the electric field is the cortex. Therefore a better comparison of TEP-to-TMP scalp electric
fields might be obtained when their respective electric fields at the radial distance of the cortex were similar.

\subsection{Comparison of TMP and TEP Power Dissipation}
\label{comp_tms_tes_power}

This section examines the ratio of TEP-to-TMP power dissipated in the scalp for similar electric fields at 
the radial distance of the cortex. That is, the quantity $R$ given by
\begin{equation}
 R = \frac{P^{tep}_2}{P^{tmp}_2} = \frac {\sum_{jm} P^{tep,2}_{jm}}{\sum_{jm} P^{tmp,2}_{jm}}
  \label{eq_2.22u} 
\end{equation}
is calculated where $P^{tep,2}_{jm}$ and $P^{tmp,2}_{jm}$ are respectively the TEP and TMP power dissipated 
in region 2 for the VSH component indexed by $(j,m)$. Such a quantity would allow one to meaningfully compare 
TEP to TMP electric fields with respect to the energy they dissipate in the scalp.

Importantly, the TEP and TMP electric fields cannot be equal since they reside in orthogonal subspaces. Given 
this limitation, a metric of the electric field similarity must be defined. The metric used here defines similar 
TEP and TMP electric fields as those which have identical cortical surface-area-averaged power dissipation for 
each component of their VSH expansion at all times $t$. This condition is insufficient to uniquely define the 
similar electric fields but it is a reasonable definition of similarity and is, as will bee seen, sufficient 
to derive $R$.

For the TEP or TMP electric field the power $P_S(\alpha_0)$ averaged over the spherical surface $S$ at 
$r=\alpha_0$ (the cortical surface) is given by 
\begin{eqnarray}  
  P_S(\alpha_0) 
  = \sigma_k \int_{S_0} |{\bf e}(\alpha_0,\theta, \phi)|^2 \sin \theta d\theta d\phi
  \label{eq_g.22z}
\end{eqnarray}
The chosen metric of similarity requires that $P^{TMP}_S(\alpha_0) = P^{TEP}_S(\alpha_0)$ and according to 
equations (\ref{eq_1.3}) and (\ref{eq_2.85}) similarity is obtained when:
\begin{eqnarray}
  \overline{\left| \frac{\partial J^j_{jm}}{\partial t} \right|^2} = \frac{a^2_j}{r_2^2 \alpha_0^2} 
  j (2j+1)^3 \alpha_0^{-(4j+2)} \alpha_1^{-(4j+2)} {\mathcal D}_j^{-2} \overline{|I_{jm}|^2}
  \label{eq_f.22r} 
\end{eqnarray} 
or alternatively
\begin{eqnarray}
  {\mathcal D}_j^{-2} \overline{|I_{jm}|^2} = \frac{r_2^2 \alpha_0^2}{a^2_j} 
  \frac{1}{j (2j+1)^3} \alpha_0^{4j+2} \alpha_1^{4j+2} 
  \overline{\left| \frac{\partial J^j_{jm}}{\partial t} \right|^2} .
  \label{eq_f.22rb} 
\end{eqnarray} 
According to equations (\ref{eq_1.24}) and (\ref{eq_2.21}) the ratio of volume-averaged power dissipated 
in the scalp (region 2) is
\begin{eqnarray}
 R 
 &=& 
 \frac
 {
   \sum_{jm} j {\mathcal D}_j^{-2} \overline{|I_{jm}|^2} 
   \left[ 
     d_{0j} \alpha_0^{-(2j+1)} \alpha_1^{-(2j+1)} + d_{1j} \alpha_1^{-(4j+2)}  
   \right]^2
   \left[
     1 - \alpha_1^{2j+1}
   \right]
 } 
 { 
   r_2^2 \sum_{jm} \overline{\left| \frac{\partial J^j_{jm}}{\partial t} \right|^2} \left[ (2j+1)^2 (2j+3) \right]^{-1} 
   \left[ 1 - \alpha_1^{2j+3} \right] 
 }
 \nonumber \\
 &+&
  \frac
 {
   \sum_{jm} (j+1) {\mathcal D}_j^{-2} \overline{|I_{jm}|^2} e_j^2 \left[ \alpha_0^{-(2j+1)} - \alpha_1^{-(2j+1)} \right]^2
   \left[
     \alpha_1^{-(2j+1)} - 1
   \right]
 } 
 {
   r_2^2 \sum_{jm} \overline{\left| \frac{\partial J^j_{jm}}{\partial t} \right|^2} \left[ (2j+1)^2 (2j+3) \right]^{-1} 
   \left[ 1 - \alpha_1^{2j+3} \right] 
 }
 \label{eq_2.22s} 
\end{eqnarray}
and substituting (\ref{eq_f.22r}) into (\ref{eq_2.22s}) yields:
\begin{eqnarray}
 R 
 &=&  
 \frac
 {
   \alpha_0^2 \sum_{jm} j  
   \left[ 
     d_{0j} \alpha_0^{-(2j+1)}\alpha_1^{-(2j+1)} + d_{1j} \alpha_1^{-(4j+2)}  
   \right]^2 
   \left[ 1 - \alpha_1^{2j+1} \right] {\mathcal D}_j^{-2} \overline{|I_{jm}|^2} 
 } 
 { 
   \sum_{jm} j(2j+1)(2j+3)^{-1} a_j^2 \alpha_0^{-(4j+2)} \alpha_1^{-(4j+2)} 
   \left[ 1 - \alpha_1^{2j+3} \right] {\mathcal D}_j^{-2} \overline{|I_{jm}|^2} 
 }
 \nonumber \\
 &+&
  \frac
 {
   \alpha_0^2 \sum_{jm} (j+1) e_j^2 \left[ \alpha_0^{-(2j+1)} - \alpha_1^{-(2j+1)} \right]^2
   \left[ \alpha_1^{-(2j+1)} - 1 \right] {\mathcal D}_j^{-2} \overline{|I_{jm}|^2} 
 } 
 {
   \sum_{jm} j(2j+1)(2j+3)^{-1} a_j^2 \alpha_0^{-(4j+2)} \alpha_1^{-(4j+2)}  
   \left[ 1 - \alpha_1^{2j+3} \right] {\mathcal D}_j^{-2} \overline{|I_{jm}|^2} 
 }
 \label{eq_f.22s} 
\end{eqnarray}
Alternatively by substituting equation (\ref{eq_f.22rb}) into (\ref{eq_2.22s}) the ratio for similar electric fields 
within the brain (region 0) becomes 
\begin{eqnarray}
 R 
 &=&  
 \frac
 {
   \alpha_0^2 \sum_{jm} a_j^{-2} (2j+1)^{-3} 
   \overline{\left| \frac{\partial J^j_{jm}}{\partial t} \right|^2}   
   [d_{0j} + d_{1j} (\alpha_0/\alpha_1)^{2j+1}]^2 
   \left[ 1 - \alpha_1^{2j+1} \right]  
 } 
 { 
   \sum_{jm} [(2j+1)^2(2j+3)]^{-1}  
   \left[ 1 - \alpha_1^{2j+3} \right] 
   \overline{\left| \frac{\partial J^j_{jm}}{\partial t} \right|^2} 
 }
 \nonumber \\
 &+&
  \frac
 {
   \alpha_0^2 \sum_{jm} (e_j/a_j)^2 \frac{j+1}{j(2j+1)^3} \left[ \alpha_1^{2j+1} - \alpha_0^{2j+1} \right]^2
   \left[ \alpha_1^{-(2j+1)} - 1 \right] 
   \overline{\left| \frac{\partial J^j_{jm}}{\partial t} \right|^2} 
 } 
 {
   \sum_{jm} [(2j+1)^2(2j+3)]^{-1}  
   \left[ 1 - \alpha_1^{2j+3} \right] \overline{\left| \frac{\partial J^j_{jm}}{\partial t} \right|^2} 
 }
 \label{eq_f.22sb} 
\end{eqnarray}
Whether to choose equation (\ref{eq_f.22s}) or (\ref{eq_f.22sb}) depends upon whether one is comparing similar 
fields generated by a given TEP electrode configuration ($I_{jm}$ are known) or by a given TMP field ($J^j_{jm}$ 
are known).

If the $I_{jm}$ are zero for all but one value of $j$ ($m$ unrestricted) then
\begin{eqnarray}
 R_j 
  &=& 
  \left( \frac{\alpha_0}{a_j} \right)^2 \frac{2j+3}{2j+1}
  \left[ 
    d_{0j} + d_{1j} (\alpha_0/\alpha_1)^{2j+1}  
  \right]^2 
  \left[ \frac{1 - \alpha_1^{2j+1}}{1 - \alpha_1^{2j+3}} \right]
  \nonumber \\
  &+& 
  \left( \frac{\alpha_0}{a_j} \right)^2 \frac{2j+3}{2j+1}
  \frac{j+1}{j} e_j^2 
  \alpha_1^{2j+1} \left[1 - (\alpha_0/\alpha_1)^{2j+1} \right]^2
  \left[ \frac{1 - \alpha_1^{2j+1}}{1 - \alpha_1^{2j+3}} \right].
 \label{eq_f.22t2} 
\end{eqnarray}
As $j \to \infty$ the value of $R_j \to R_{\infty}$ where 
\begin{equation}
   R_{\infty} = \frac{1}{16} \frac{(1+\epsilon)^4}{\epsilon^2} \alpha_0^2
  \label{eq_f.22s2}
\end{equation}
{\it That is, $R_j$ asymptotically approaches an upper limit determined by the relative size of the brain region $\alpha_0$ 
and the scalp-to-skull conductivity ratio $\epsilon$}. Note that if the TEP and TMP electric fields are constrained to be similar at arbitrary depth, rather than at the cortical surface, then $\alpha_o$ is replaced by $\alpha = r/r_2$. Therefore the advantage of TMP over TEP with respect to the scalp-to-brain ratio of the root-mean-squared electric field is linear with respect to radial depth at which the fields are taken to be similar.

\begin{figure}[ht]
    \input{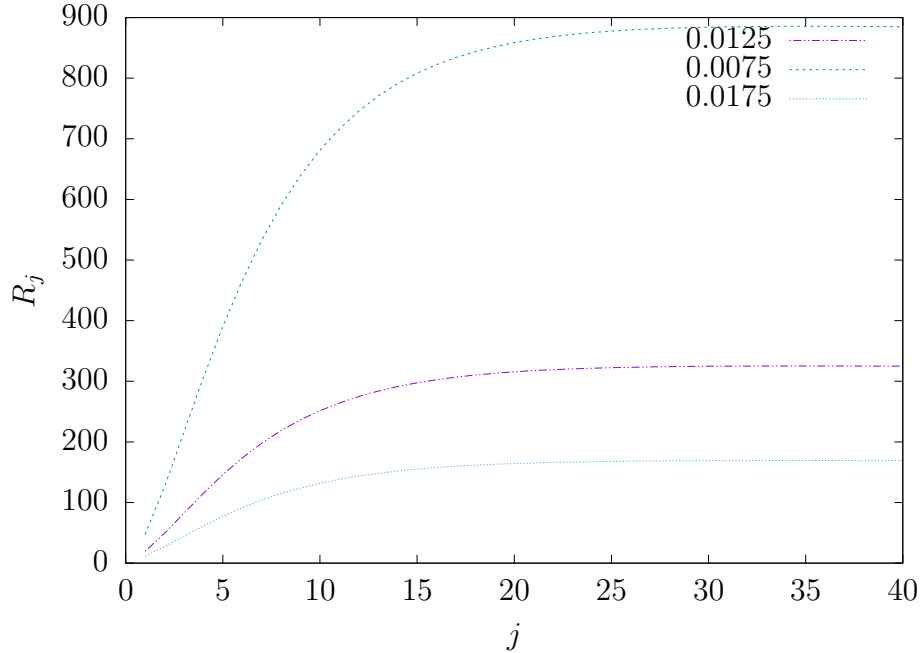}
    \caption{The ratio $R_j$ versus $j$ for three different values of the skull-to-scalp conductivity 
    ratio $\epsilon$. $R_j$ is the ratio of the TEP to TMP mean-squared electric field over the scalp for similar electric fields at the level of the cortical surface when the electric field is comprised of a single vector spherical harmonic component indexed by integers $j$ and $m$. The literature typically uses the value $\epsilon = 1/80 = 0.0125$. Note that each curve approaches an asymptotic limit given by $R_{\infty} = \frac{1}{16} \frac{(1+\epsilon)^4}{\epsilon^2} \alpha_0^2$. For similar electric fields at the level of the cortex, TEP produces much larger mean-squared scalp electric fields compared to TMP as electric field focality increases at the cortex.}
    \label{R_j_plot}
\end{figure}

Figure \ref{R_j_plot} gives the plot of $R_j$ versus $j$ for three different conductivity ratios $\epsilon$ although
$\epsilon= 0.0125$ is the value most often assumed in the literature. The plot shows that $R_j$ clearly increases 
with respect to $j$, an index of decreasing spatial scale of the electric field, but reaches an asymptotic value. 
For the particular dimensions of three-shell model used in this work the asymptotic values corresponding to the 
conductivity ratios $\epsilon = 0.0075, 0.0125$ and $0.0175$ are $865.6, 317.9$ and $165.4$. 

\begin{figure}[ht]
    \input{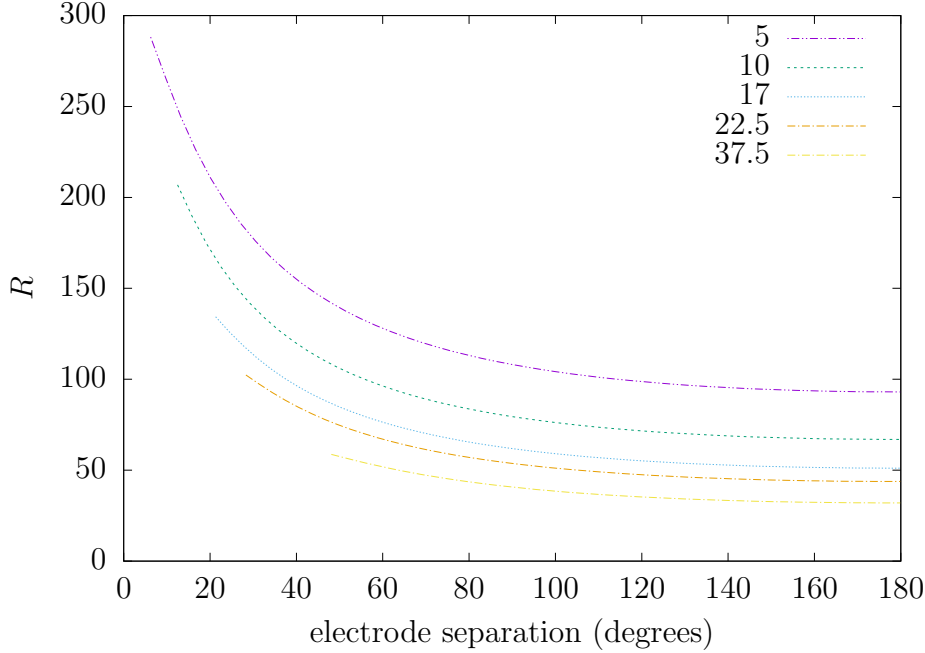}
    \caption{$R$ versus the electrode separation angle (degrees) for five different electrode radii ($\rm{mm}$)
    that are typical of commercial TEP electrode systems depicted in figure \ref{tep_electrodes}. $R$ is the ratio of the TEP to TMP mean-squared electric field over the scalp when the TMP electric field at the level of the cortical surface is similar to that produced by the TEP system depicted in figure \ref{tep_electrodes}. Each plot extends from the minimal separation between electrodes 
    (twice the angle subtended by the electrode from its center) up to 180$^\circ$.}
    \label{R_TES_plot}
\end{figure}

\begin{figure}[ht]
    \input{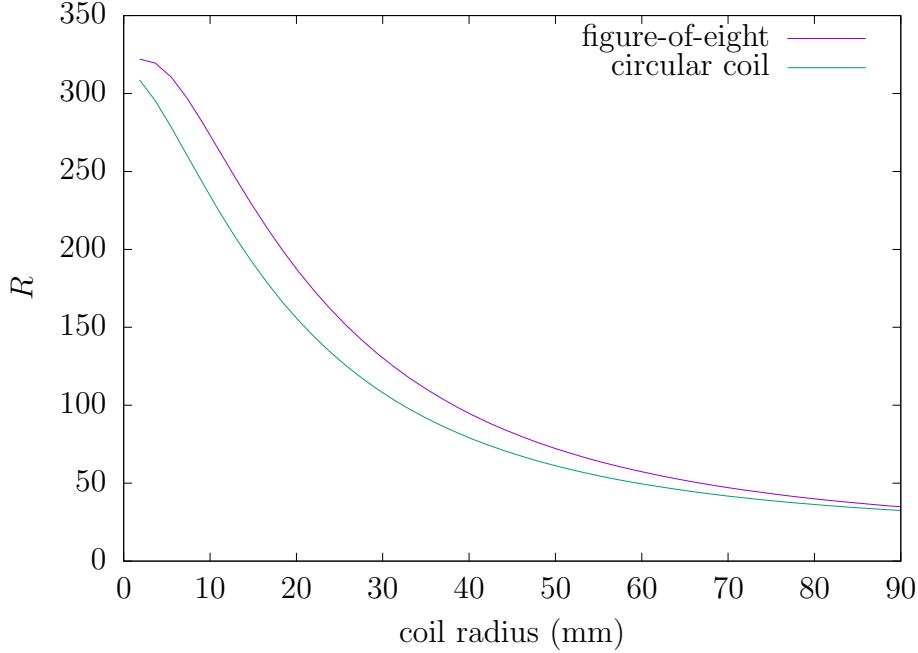}
    \caption{$R$ versus coil radius of the circular and figure-8 coils ($\epsilon =$ 1/80). $R$ is the ratio of the TEP to TMP mean-squared electric field over the scalp when the TEP electric field at the level of the cortical surface is similar to that produced by the TMP system depicted in figure \ref{tmp_coil}.} 
    \label{R_TMP_plot}
\end{figure}

Figure \ref{R_TES_plot} gives a plot of $R$ versus electrode separation angle for five different electrode radii
typical in TEP systems. As the electrode separation or the electrode radii decrease --- that is, the spatial detail
in the field increases --- the value of $R$ increases.  Figure \ref{R_TMP_plot} gives a plot of $R$ versus coil 
radius for the simple circular TMS coil and the figure-8 coil. Again note that as the spatial detail 
increases (coil size decreases) $R$ increases. For typical circular coils with 
average radii of 20 to 40 mm the value of corresponding values of $R$ range from 105 to 171. Also note that, as 
compared to the circular coil of the same radius, the figure-8 coil has a modestly increased value of $R$. 

Note, by referring to equation (\ref{eq_2.22u}), that $R_{\infty}$, the asymptotic value of $R_j$, is additionally an 
upper bound to the value of $R$. Therefore although $R$, the TEP-to-TMP ratio of power dissipated in the scalp 
for similar electric fields at the cortex, increases with increasing spatial detail of the field there is an upper 
limit to this value. Also note that the electric field of a figure-8 coil, which is in common use in TMS research, 
would be expected to yield values of $R$ considerably greater than that of the circular coil used here due to its more
focal field.

\subsection{TMP and TEP Power Dissipation for Nonseparable Current Densities}
\label{nonsep}

All of the current densities considered in this work have, up to this point, been separable with respect to time and spatial coordinates as is typical of extant TEP and TMP systems. However, interesting TEP work has recently been done with nonseparable current densities to produce spatially dependent temporal interference effects in mice brains \cite{Boyden}. Consider now the nonseparable case of a TMP system comprised of the two circular coils of the figure-8 example albeit with each coil driven independently. Coil 1 has time dependence $\sin \omega_1 t$ and coil 2 has time dependence $\sin \omega_2 t$. In such a case:
\begin{eqnarray}
  \sum_m \left| 
    \frac{\partial J^{j}_{jm}}{\partial t} 
  \right|^2 
  &=& 
  \sum_m \left|
    \omega_1 \cos \omega_1 t J^{j+}_{jm} - \omega_2 \cos \omega_2 t J^{j-}_{jm}
  \right|^2
  \nonumber \\
  &=&
  \omega_1^2 \cos^2 \omega_1 t \sum_m |J^{j+}_{jm}|^2 + \omega_2^2 \cos^2 \omega_2 t \sum_m |J^{j-}_{jm}|^2 
  \nonumber \\
  &-& 
  \omega_1 \omega_2 \cos \omega_1 t \cos \omega_2 t \sum_m (J^{j+*}_{jm} J^{j-}_{jm} + J^{j+}_{jm} J^{j-*}_{jm})
  \nonumber \\
  &=&
  \omega_1^2 \cos^2 \omega_1 t \sum_m |J^{j+}_{jm}|^2 + \omega_2^2 \cos^2 \omega_2 t \sum_m |J^{j-}_{jm}|^2 
  \nonumber \\
  &-& 
  \frac{\omega_1 \omega_2}{2} [\cos (\omega_1 - \omega_2)t + \cos (\omega_1 + \omega_2)t] 
  \sum_m [J^{j+*}_{jm} J^{j-}_{jm} + J^{j+}_{jm} J^{j-*}_{jm}]
  \label{eq_f.22sb103.a}
\end{eqnarray}
where $J^{j+}_{jm}$ and $J^{j-}_{jm}$ are the coefficients corresponding to coils 1 and 2 (see Appendix \ref{J_coef}).
For time averages over an interval $T$ such $|\omega_1 - \omega_2|^{-1} \ll T$
\begin{eqnarray}
  \sum_m \overline{
   \left| 
    \frac{\partial J^{j}_{jm}}{\partial t} 
  \right|^2} 
  &\approx&
  \frac{1}{2} \left( \omega_1^2 \sum_m |J^{j+}_{jm}|^2 + \omega_2^2 \sum_m |J^{j-}_{jm}|^2 \right)
  \label{eq_f.22sb103.b}
\end{eqnarray}
where the horizontal line denotes a time average. Finally, since the coils are assumed to be identical in their shape 
\begin{eqnarray}
  \sum_m \overline{
   \left| 
    \frac{\partial J^{j}_{jm}}{\partial t} 
  \right|^2} 
  &\approx&
  \frac{1}{2} \left( \omega_1^2 + \omega_2^2 \right) \sum_m |J^{j+}_{jm}|^2
  \label{eq_f.22sb103.b1}
\end{eqnarray}
For the TEP case of two pairs of electrodes, each pair driven by current sources of different frequencies, a similar result is obtained: 
\begin{eqnarray}
  \sum_m \overline{| I_{jm} |^2} 
  &\approx&
  \sum_m |I^{+}_{jm}|^2.
  \label{eq_f.22sb103.c}
\end{eqnarray}
Therefore in the long-time average case a system of two pairs of TEP electrodes, in which each pair is identical except for a rotation on the sphere's surface, the power ratios calculated for the interfering pair are the same as that for a single pair of electrodes. A similar statement can be made for interfering TMP coils except that the single coil power ratio is multiplied by the average of the squared frequencies of each coil. These results can be extrapolated to an arbitrary number of coils or electrode pairs rotated to different positions on a spherical surface in which the long-time average extends over an time interval large compared to the reciprocal of the smallest frequency differences.

\subsection{Current Source Energy Utilization}
\label{energy}

From the results given in the previous sections it is clear that TMP has the distinct advantage of producing 
a much smaller scalp electric field than TEP for similar cortical fields and therefore capable of diminishing 
potential deleterious scalp effects. However many TEP applications target specific brain electric field frequencies in the range of 0 - 200 Hz. This frequency range covers most of the brain frequencies which are measured by 
electroencephalography and magnetoencephalography. As the simple analysis of this section will show, within this 
frequency range it is energetically costly to generate TMP electric fields of significant amplitude to potentially 
alter brain activity (at least 0.5 V/m).

The source of electric current for TEP or TMP delivers power to a load which is comprised of cables, electrodes 
(TEP) or coils (TMP) and a head. For standard TES and TMS systems there is one source of current driving the TES 
electrodes or the TMS coil. Because of this the quasistatic electric field ${\bf E}({\bf r},t)$ induced in the 
brain will be separable with respect to temporal and spatial variables. As already noted the TES electric field 
depends linearly upon the current amplitude $I_e(t)$ whereas the TMS electric field depends linearly upon the 
temporal derivative of the current  $I_m(t)$ supplied to the coil. In the following it will be assumed that
$I_m(t)=I_{mo}\sin(2\pi ft)$ and $I_e(t)=I_{eo}\sin(2\pi ft)$ where $f$ is the frequency of a continuous applied 
field. We can then write the corresponding electric fields as
\begin{eqnarray}
  {\bf E}_m({\bf r},t) =  e_m \; \frac{dI_m(t)}{dt} \; {\bf e}_m({\bf r}) 
  = e_m 2\pi f I_{mo} \cos(2\pi ft) \; {\bf e}_m({\bf r}) 
  \label{eq_90.1b2}
  \\
  {\bf E}_e({\bf r},t) =  e_e \; I_e(t) \; {\bf e}_e({\bf r}) = e_e I_{eo} \sin(2\pi ft) \; {\bf e}_e({\bf r}) 
  \qquad 
  \label{eq_90.1b}
\end{eqnarray}
where ${\bf e}_m({\bf r})$ and ${\bf e}_e({\bf r})$ are vector fields with magnitude normalized to 
one at some point ${\bf r}_o$ in the cortex, and $e_m$ and $e_e$ are the corresponding magnitudes 
at that point in units of (Vs/Am) and (V/Am) respectively.

Typical TMS coils (figure-8 shape with inductance of 12.0 $\mu$H and resistance of 12.0 m$\Omega$) 
are known to produce a peak electric field amplitude of approximately 100 V/m electric in the cortex near the 
coil (ie ${\bf r}_o$) when $I_{mo} =$ 5.0 kA and $\nu_e =$ 4kHz. Using equation (\ref{eq_90.1b2}) an estimate of 
$e_m = ({\rm 100 V/m})/(2\pi I_{mo} \nu_m) = 7.9 \times 10^{-7} {\rm Vs/Am}$ is obtained. For 
a typical TES system the peak electric field is known to be approximately 0.5 V/m for $I_{eo} =$ 2.0 mA. This 
yields an estimate of $e_e = ({\rm 0.5 V/m})/I_{eo} = {\rm 250 V/Am}$.

To obtain estimates of the power supplied to the TES and TMS loads it is assumed that typical TES and TMS 
systems are used to create electric fields which have equal amplitudes at some some point ${\bf r}_o$ in 
the brain region. The point ${\bf r}_o$ will be assumed to be a relative spatial maximum (true extrema
cannot exist) for both the TEP and TMP electric fields but the distribution of the electric field about 
${\bf r}_o$ will be assumed to be only as similar as present methods allow. The ratio of temporally-averaged 
power (averaged over one period of a sinusoidal source of frequency $\nu_m$) supplied by the TES or TMS current 
source to the respective loads can be written as:
\begin{eqnarray}  
  r  
  = \frac{I^2_{eo} R_e}{I^2_{mo} R_m} 
  \label{eq_151a}
\end{eqnarray}
where $R_e$ and $R_m$ are the resistances of the TEP and TMP loads respectively. The TES load is primarily due 
to the resistance at the scalp-electrode interface and to lesser degree on the conductivity of the head and cables. 
The resistance of the TMS load is primarily due the resistance of the TMS coil and cable. We have previously noted that
$I_{eo} = |{\bf E}_e|/e_e$ and $I_{mo} = |{\bf E}_m|/(2\pi e_m \nu_m)$ where $|{\bf E}_e|$ and $|{\bf E}_m|$ are values
for the fields at $r_o$. Therefore we can write:
\begin{eqnarray}  
  r  
  = \left( \frac{2\pi \nu_m e_m}{e_e} \right)^2  
  \left( \frac{|{\bf E}_e|}{|{\bf E}_m|} \right)^2 
  \frac{R_e}{R_m}
  \label{eq_151b}
\end{eqnarray}
Since the electric field amplitudes are assumed to be equal at ${\bf r}_o$ we can write
\begin{eqnarray}  
  r  
  = \left( \frac{2\pi \nu_m e_m}{e_e} \right)^2  
  \frac{R_e}{R_m}
  \label{eq_151c}
\end{eqnarray}
Since $R_e$ is primarily due to the scalp-electrode interface it is roughly independent of the position of the TES 
electrodes. Also $R_m$ is roughly independent of the presence of the head. Reasonable estimates for the two 
quantities are $R_e = {\rm 10 } k\Omega$ and $R_m = {\rm 12} m\Omega$. These estimates correspond to those given 
for NeuroConn TES electrodes and a figure-eight MagVenture TMS coil. If we insert the values for $e_e$ and $e_m$ 
(determined in the previous section) for the human head along with resistances $R_e$ and $R_m$ of typical human head 
systems we obtain
\begin{eqnarray}  
  r \approx \nu_m^2 ({\rm 3.3} \times {\rm 10}^{-10})  
  \label{eq_151d}
\end{eqnarray}
Notice that this estimate depends on the square of the frequency. For a frequency of 10 Hz we have $r \approx
{\rm 3.3} \times {\rm 10}^{-8}$ whereas for a frequency of 55 kHz we have $r \approx 1.0$. {\it Clearly TEP is far 
more energy efficient than TMP at low frequencies whereas the reverse is true at very high frequencies}. To obtain a 10Hz 0.5 V/m TES electric field amplitude within the 
cortex requires the current source to supply approximately 2.0 mA to an electrode pair. The average power per 
cycle is then $\rm{0.5} \times \rm{(2.0 mA)^2} \times {\rm 10 } k\Omega = \rm{0.02 W}$. Using equation 
(\ref{eq_151d}) we can estimate that achieving the same TMP electric field using a typical human TMS coil would 
require approximately 610 kW.

\section{Discussion}
\label{discuss}

It is well known that TEP amd TMP electric fields cannot have extremal points within the interior of the head. The 
extremal points must always occur at the boundaries hence the scalp electric field will always be of greater magnitude 
than the brain electric field regardless of the method. As scalp electric fields increase in magnitude they may elicit 
pain due to coupling with peripheral nerves. At higher magnitudes still potentially dangerous effects due to scalp 
heating may occur. These deleterious effects set a maximum electric field magnitude within the scalp and consequently 
within the brain. However with low frequency TMP (e.g. 0-200 Hz) it is energetically costly to generate electric 
fields of sufficient magnitude to significantly influence neuronal state. Understanding how electric field focality, 
scalp heating and energy utilization shape the experimental TEP and TMP space is of value to the researcher and 
inventor of new noninvasive brain perturbation methods and technology.

In this work the analytic solutions of the TEP and TMP three shell model are derived and used to demonstrate 
important features of the respective electric fields and to estimate scalp-to-brain mean-square electric field
ratios as well as the TEP-to-TMP ratio scalp mean-squared electric fields for similar electric fields at the cortex. 
When looking for general principles and model-based estimates analytic solutions are superior to numerical solutions 
since they obtain a general solution based on system variables rather a set of specific solutions based on specific 
choices of variables. Of the general features elucidated here:

\begin{enumerate}
  \item TEP and TMP electric fields exist in orthogonal subspaces spanned by the vector spherical 
        harmonics ${\bf Y}_{jm}^{j-1}(\theta, \phi)$ and ${\bf Y}_{jm}^j(\theta, \phi)$ respectively. The 
        ${\bf Y}_{jm}^j(\theta, \phi)$ vector spherical harmonics have no radial component whereas the 
        ${\bf Y}_{jm}^{j-1}(\theta, \phi)$ do. Therefore a TMP electric field can only be tangential to the surface of
        the spherical head (as has been noted elsewhere \cite{WangPeterchev}).

  \item TEP and TMP can have similar focality in the absence of the restrictions set by scalp mean-square electric field. 
        A given value of index $j$ adds similar levels of angular spatial detail on a sphere of arbitrary radius within 
        the head for both TEP and TMP electric fields. 
 
  \item For both methods as the angular spatial detail (indexed by $j$) in the electric field increases so does the ratio 
        of power dissipated in the scalp (or mean square scalp electric field) to that in the brain. For typical conductance
        values of the three-shell human head model this ratio is much higher in TEP ($R^{tep}_j$) than TMP ($R^{tmp}_j$). 

  \item For similar electric fields at the radial distance of the cortex there exists an upper bound to the ratio of 
        TEP-to-TMP mean-square scalp electric field given by the quantity $R_{\infty}$. A value of approximately 318 
        was calculated for typical human head three-shell model parameters. Note that the root-mean-square electric field ratio would therefore be 17.8.

  \item At low frequencies (0-200 Hz) the energetic cost for a current source to generate electric fields of appreciable
        magnitude within the brain region are much higher for TMP as compared to TEP.

\end{enumerate}

The energetic cost associated with TMP could be made practical if electric fields of frequency greater than 
1 kHz were used to perturb brain function. Recent publications suggest that this may be possible. It is
well known that suprathreshold electric fields are able to robustly produce electrical nerve block in peripheral
nerves \cite{KilBha1}. It has recently been shown that amplitude modulation of suprathreshold kilohertz frequency
TEP electric fields \cite{Boyden} may allow some degree of spatial localization with respect to the radial variable 
$r$ by means of spatially distributed interference effects. The proposed mechanism is such that the amplitude
of the modulation, rather than the amplitude and frequency of the electric field alone, plays a role in the 
coupling to neurons. The amplitude of the modulation can vary spatially thereby allowing additional spatial 
localization of effects in a manner not restricted by the extremum principle. Furthermore subthreshold TEP at 
2-5 kHz and 2.0 mA has been shown to effect motor evoked potentials with approximately the same efficacy as TEP 
in the 0-100 Hz range \cite{Chaieb}. These results suggest that continuously applied kHz TMP electric fields may 
be an effective and energetically feasible method to perturb brain states and function.

It should be noted that if kHz amplitude modulation does play a role in spatial focusing of TEP electric fields then 
this method could allow one to increase the spatial localization of electric field effects without increasing the
mean-squared electric field within the scalp. Although this would be a welcome finding, kHz TMP amplitude modulation
methods could increase the localization or amplitude of brain electric fields amplitude obtained from kHz even further.
However this increase would come at a cost since, as has been shown, TEP systems are less energetically costly as
compared to TMP systems.

\appendix
\appendixpage
\section{Vector Spherical Harmonic Definitions and Properties}
\label{VSH_props}

The $l=j-1,j,j+1$ VSH components are defined as follows:
\begin{eqnarray}
  {\bf Y}_{jm}^{j+1}(\theta,\phi) 
  &=& \sqrt{\frac{j+1}{2j+1}} \left(-{\bf e}_{r} Y_{jm}(\theta,\phi) 
  + {\bf e}_{\theta} \frac{1}{j+1} \frac{\partial Y_{jm}(\theta,\phi)}{\partial \theta} 
  + {\bf e}_{\phi} \frac{im}{j+1} \frac{Y_{jm}(\theta,\phi)}{\sin \theta} \right)
  \nonumber \\
  {\bf Y}_{jm}^{j}(\theta,\phi) 
  &=& -{\bf e}_{\theta} \frac{m}{\sqrt{j(j+1)}} \frac{Y_{jm}(\theta,\phi)}{\sin \theta} 
  - {\bf e}_{\phi} \frac{i}{\sqrt{j(j+1)}} \frac{\partial Y_{jm}(\theta,\phi)}{\partial \theta}
  \nonumber \\
  {\bf Y}_{jm}^{j-1}(\theta,\phi) 
  &=& \sqrt{\frac{j}{2j+1}} \left({\bf e}_{r} Y_{jm}(\theta,\phi) 
  + {\bf e}_{\theta} \frac{1}{j} \frac{\partial Y_{jm}(\theta,\phi)}{\partial \theta} 
  + {\bf e}_{\phi} \frac{im}{j} \frac{Y_{jm}(\theta,\phi)}{\sin \theta} \right)
  \label{app_0} 
\end{eqnarray}
The VSH components have many interesting properties. The following properties will be of use in the derivations 
presented in this work: 
\begin{eqnarray}
  {\hat{\bf r}} \cdot {\bf Y}^{j+1}_{jm}(\theta,\phi) &=& 
  - \left(\frac{j+1}{2j+1}\right)^{1/2} Y_{jm}(\theta,\phi)
  \nonumber \\
  {\hat{\bf r}} \cdot {\bf Y}^{j}_{jm}(\theta,\phi) &=& 0
  \nonumber \\
  {\hat{\bf r}} \cdot {\bf Y}^{j-1}_{jm}(\theta,\phi) &=& 
  \left(\frac{j}{2j+1}\right)^{1/2} Y_{jm}(\theta,\phi)
  \label{app_1} 
\end{eqnarray}
\begin{eqnarray}
  {\hat{\bf r}} \times {\bf Y}^{j+1}_{jm}(\theta,\phi) &=& 
  i \left(\frac{j}{2j+1}\right)^{1/2} {\bf Y}^{j}_{jm}(\theta,\phi)
  \nonumber \\
  {\hat{\bf r}} \times {\bf Y}^{j}_{jm}(\theta,\phi) &=& 
  i \left(\frac{j+1}{2j+1}\right)^{1/2} {\bf Y}^{j-1}_{jm}(\theta,\phi)
  + i \left(\frac{j}{2j+1}\right)^{1/2} {\bf Y}^{j+1}_{jm}(\theta,\phi)
  \nonumber \\
  {\hat{\bf r}} \times {\bf Y}^{j-1}_{jm}(\theta,\phi) &=&  
  i \left(\frac{j+1}{2j+1}\right)^{1/2} {\bf Y}^{j}_{jm}(\theta,\phi)
  \label{app_2} 
\end{eqnarray}

\section{Solving for the TEP E Field}
\label{solv_tes}

We will assume that $\sigma_2 = \sigma_0$ and write $\epsilon = \sigma_1/\sigma_0$. In all cases of interest 
the conductivity of the skull will be much less than the conductivity of the scalp or brain and therefore 
$\epsilon << 1$. For our purpose we will use the usual ratio of $\epsilon = 1/80 = 0.0125$. Applying the 
boundary condition at $r=1$ we have
\begin{eqnarray}
  {\bf j}(1,\theta,\phi) \cdot {\hat{\bf r}}
  &=& 
  - \sigma_2 \sum_{jm} 
  [j(2j+1)]^{1/2} D_{jm} {\bf Y}^{j-1}_{jm}(\theta,\phi) \cdot {\hat{\bf r}} 
  \nonumber \\
  &-&
  \sigma_2 \sum_{jm}
  [(j+1)(2j+1)]^{1/2} E_{jm} {\bf Y}^{j+1}_{jm}(\theta,\phi) \cdot {\hat{\bf r}} 
  \nonumber \\
  &=& - \sigma_2 \sum_{jm} 
  \left[
    j D_{jm} - (j+1) E_{jm} 
  \right]
  Y_{jm}(\theta,\phi) 
  \label{eq_A1.7}
\end{eqnarray}
and using the VSH orthogonality relationship we get
\begin{equation}
  - j D_{jm} + (j+1) E_{jm} = I_{jm}  
\label{eq_A1.8}
\end{equation}
where
\begin{equation}
  I_{jm} = \frac{1}{\sigma_2} 
  \int_0^{2\pi} \int_0^{\pi} {\bf j}(1,\theta,\phi) \cdot {\hat{\bf r}} \; Y^*_{jm}(\theta,\phi) 
  \sin \theta d\theta d\phi 
\label{eq_A1.9}
\end{equation}
Note that since $\nabla \cdot {\bf j} = 0$ for a quasistatic system then, according to Gauss's Law, 
$I_{00} = 0$ therefore the double summation indices are now $j=1, \ldots, \infty$ and $m=-j, \ldots, j$.

Applying the first boundary condition at $r=\alpha_1$ we have
\begin{eqnarray}
  & & \sigma_1 \sum_{jm} 
  [j(2j+1)]^{1/2} B_{jm} \alpha_1^{j-1} {\bf Y}^{j-1}_{jm}(\theta,\phi)  \cdot {\hat{\bf r}} 
  \nonumber \\
  &+& 
  \sigma_1 \sum_{jm} 
  [(j+1)(2j+1)]^{1/2} C_{jm} \alpha_1^{-(j+2)} {\bf Y}^{j+1}_{jm}(\theta,\phi)  \cdot {\hat{\bf r}} 
  \nonumber \\ 
  &=& 
  \sigma_0 \sum_{jm} 
  [j(2j+1)]^{1/2} D_{jm} \alpha_1^{j-1} {\bf Y}^{j-1}_{jm}(\theta,\phi)  \cdot {\hat{\bf r}}
  \nonumber \\ 
  &+& 
  \sigma_0 \sum_{jm} 
  [(j+1)(2j+1)]^{1/2} E_{jm} \alpha_1^{-(j+2)} {\bf Y}^{j+1}_{jm}(\theta,\phi)  \cdot {\hat{\bf r}} 
  \label{eq_A1.10.1a}
\end{eqnarray}
or
\begin{eqnarray}
  \sigma_1 \sum_{jm}
  \left[
    j B_{jm} \alpha_1^{j-1} - (j+1) C_{jm} \alpha_1^{-(j+2)} 
  \right] Y_{jm}(\theta,\phi) 
  \nonumber \\ 
  = \sigma_0 \sum_{jm}
  \left[
    j D_{jm} \alpha_1^{j-1} - (j+1) E_{jm} \alpha_1^{-(j+2)} 
  \right] Y_{jm}(\theta,\phi) 
\label{eq_A1.10}
\end{eqnarray}
therefore
\begin{equation}
  \epsilon j B_{jm} - \epsilon (j+1) C_{jm} \alpha_1^{-(2j+1)} 
  - j D_{jm} + (j+1) E_{jm} \alpha_1^{-(2j+1)} = 0
\label{eq_A1.11}
\end{equation}

Applying the second boundary condition at $r=\alpha_1$ we have
\begin{eqnarray}
  & &\sum_{jm} 
  [j(2j+1)]^{1/2} B_{jm} \alpha_1^{j-1} {\hat{\bf r}} \times {\bf Y}^{j-1}_{jm}(\theta,\phi) 
  \nonumber \\ 
  &+& 
  \sum_{jm} 
  [(j+1)(2j+1)]^{1/2} C_{jm} \alpha_1^{-(j+2)} {\hat{\bf r}} \times {\bf Y}^{j+1}_{jm}(\theta,\phi) 
  \nonumber \\ 
  &=& 
  \sum_{jm} 
  [j(2j+1)]^{1/2} D_{jm} \alpha_1^{j-1} {\hat{\bf r}} \times {\bf Y}^{j-1}_{jm}(\theta,\phi)  
  \nonumber \\
  &+& 
  \sum_{jm} 
  [(j+1)(2j+1)]^{1/2} E_{jm} \alpha_1^{-(j+2)} {\hat{\bf r}} \times {\bf Y}^{j+1}_{jm}(\theta,\phi) 
  \label{eq_A1.12d}
\end{eqnarray}
or
\begin{eqnarray}
  \sum_{jm}
  [j(j+1)]^{1/2} B_{jm} \alpha_1^{j-1} {\bf Y}^j_{jm}(\theta,\phi) 
  + \sum_{jm} 
  [j(j+1)]^{1/2} C_{jm} \alpha_1^{-(j+2)} {\bf Y}^j_{jm}(\theta,\phi)  
  \nonumber \\  
  = \sum_{jm} 
  [j(j+1)]^{1/2} D_{jm} \alpha_1^{j-1} {\bf Y}^j_{jm}(\theta,\phi) 
  + \sum_{jm} 
  [j(j+1)]^{1/2} E_{jm} \alpha_1^{-(j+2)} {\bf Y}^j_{jm}(\theta,\phi) 
  \label{eq_A1.12}
\end{eqnarray}
which yields
\begin{equation}
  B_{jm} + C_{jm} \alpha_1^{-(2j+1)} - D_{jm} - E_{jm} \alpha_1^{-(2j+1)} = 0
\label{eq_A1.13}
\end{equation}

Applying the first boundary condition at $r=\alpha_0$ we have
\begin{eqnarray}
  & &\sigma_0 \sum_{jm} 
  [j(2j+1)]^{1/2} A_{jm} \alpha_0^{j-1} {\bf Y}^{j-1}_{jm}(\theta,\phi)  \cdot {\hat{\bf r}}
  \nonumber \\ 
  &=& \sigma_1 \sum_{jm} 
  [j(2j+1)]^{1/2} B_{jm} \alpha_0^{j-1} {\bf Y}^{j-1}_{jm}(\theta,\phi)  \cdot {\hat{\bf r}} 
  \nonumber \\  
  &+&  \sigma_1 \sum_{jm} 
  [(j+1)(2j+1)]^{1/2} C_{jm} \alpha_0^{-(j+2)} {\bf Y}^{j+1}_{jm}(\theta,\phi)  \cdot {\hat{\bf r}} 
\label{eq_A1.10b.1}
\end{eqnarray}
or
\begin{eqnarray}
  \sigma_0 \sum_{jm} j A_{jm} \alpha_0^{j-1} Y_{jm}(\theta,\phi)  
  = \sigma_1 \sum_{jm}
  \left[
    j B_{jm} r_0^{j-1} - (j+1) C_{jm} \alpha_0^{-(j+2)} 
  \right] Y_{jm}(\theta,\phi) 
\label{eq_A1.10b}
\end{eqnarray}
therefore
\begin{equation}
  j A_{jm} - \epsilon j B_{jm} + \epsilon (j+1) C_{jm} \alpha_0^{-(2j+1)} = 0 .
\label{eq_A1.11b}
\end{equation}

Applying the second boundary condition at $r=\alpha_0$ we have
\begin{eqnarray}
  & & \sum_{jm} 
  [j(2j+1)]^{1/2} A_{jm} \alpha_0^{j-1} {\hat{\bf r}} \times {\bf Y}^{j-1}_{jm}(\theta,\phi)  
  \nonumber \\    
  &=& \sum_{jm} 
  [j(2j+1)]^{1/2} B_{jm} \alpha_0^{j-1} {\hat{\bf r}} \times {\bf Y}^{j-1}_{jm}(\theta,\phi) 
  \nonumber \\  
  &+& \sum_{jm} 
  [(j+1)(2j+1)]^{1/2} C_{jm} \alpha_0^{-(j+2)} {\hat{\bf r}} \times {\bf Y}^{j+1}_{jm}(\theta,\phi) 
  \label{eq_A1.12b.2}
\end{eqnarray}
or
\begin{eqnarray}
  \sum_{jm} [j(j+1)]^{1/2} A_{jm} \alpha_0^{j-1} {\bf Y}^j_{jm}(\theta,\phi)   
  &=& \sum_{jm} 
  [j(j+1)]^{1/2} B_{jm} \alpha_0^{j-1} {\bf Y}^j_{jm}(\theta,\phi) 
  \nonumber \\  
  &+& \sum_{jm} 
  [j(j+1)]^{1/2} C_{jm} \alpha_0^{-(j+2)} {\bf Y}^j_{jm}(\theta,\phi) 
  \label{eq_A1.12b}
\end{eqnarray}
which yields
\begin{equation}
  A_{jm} - B_{jm} - C_{jm} \alpha_0^{-(2j+1)} = 0. 
\label{eq_A1.13b}
\end{equation}

We will write equations (\ref{eq_A1.8}), (\ref{eq_A1.11}), (\ref{eq_A1.13}), (\ref{eq_A1.11b}) and 
(\ref{eq_A1.13b}) in the following compact form 
\begin{eqnarray}
  \begin{bmatrix} 
   0 & 0 & 0 & d^{j}_1 &e^{j}_1  \\
   0 & b^{j}_2 & c^{j}_2 & d^{j}_2 & e^{j}_2 \\
   0 & b^{j}_3 & c^{j}_3 & d^{j}_3 & e^{j}_3 \\
   a^{j}_4 & b^{j}_4 & c^{j}_4 & 0 & 0 \\
   a^{j}_5 & b^{j}_5 & c^{j}_5 & 0 & 0
  \end{bmatrix} 
    \begin{bmatrix} 
   A_{jm}  \\
   B_{jm}  \\
   C_{jm}  \\
   D_{jm}  \\
   E_{jm} 
  \end{bmatrix} 
   =
  \begin{bmatrix} 
   I_{jm} \\
   0 \\
   0 \\
   0 \\
   0
  \end{bmatrix} 
\label{eq_A1.14}
\end{eqnarray}
where
\begin{eqnarray}
 \begin{tabular}{ l l l l l}
   &  &  & $d^{j}_1 = -j $ & $e^{j}_1 = (j+1)$ \\
   & $b^{j}_2 = \epsilon j$ & $c^{j}_2 = -\epsilon (j+1) \alpha_1^{-(2j+1)}$ & $d^{j}_2 = -j$ & $e^{j}_2 = (j+1) \alpha_1^{-(2j+1)}$ \\
   & $b^{j}_3 = 1$ & $c^{j}_3 = \alpha_1^{-(2j+1)} $ & $d^{j}_3 = -1 $ & $e^{j}_3 = -\alpha_1^{-(2j+1)}$ \\
   $a^{j}_4 = j $ & $b^{j}_4 = -\epsilon j$ & $c^{j}_4 = \epsilon (j+1) \alpha_0^{-(2j+1)}$ &  & \\
   $a^{j}_5 = 1$ & $b^{j}_5 = -1$ & $c^{j}_5 = -\alpha_0^{-(2j+1)}$ &  & \\
 \end{tabular}
\label{eq_A1.14b}
\end{eqnarray}

Using Cramer's Rule the solution of this simultaneous set of equations is:
\begin{eqnarray}
  A_{jm} &=& (d^j_2 e^j_3 - d^j_3 e^j_2) (b^j_4 c^j_5 - b^j_5 c^j_4) 
            I_{jm}{\mathcal D}_j^{-1} 
  \nonumber \\
  B_{jm} &=& - (d^j_2 e^j_3 - d^j_3 e^j_2) (a^j_4 c^j_5 - a^j_5 c^j_4) 
            I_{jm}{\mathcal D}_j^{-1}
  \nonumber \\
  C_{jm} &=& (d^j_2 e^j_3 - d^{j}_3 e^j_2) (a^j_4 b^j_5 - a^j_5 b^j_4) 
            I_{jm}{\mathcal D}_j^{-1} 
  \nonumber \\
  D_{jm} &=& [(b^j_2 e^j_3 - b^j_3 e^j_2) (a^j_4 c^j_5 - a^j_5 c^j_4)
            - (c^j_2 e^j_3 - c^j_3 e^j_2) (a^j_4 b^j_5 - a^j_5 b^j_4)] 
            I_{jm}{\mathcal D}_j^{-1}   
  \nonumber \\
  E_{jm} &=& [(c^j_2 d^j_3 - c^j_3 d^j_2)(a^j_4 b^j_5 - a^j_5 b^j_4) 
	    + (b^j_3 d^j_2 - b^j_2 d^j_3) (a^j_4 c^j_5 - a^j_5 c^j_4)]
            I_{jm}{\mathcal D}_j^{-1} 
  \label{eq_A1.15a}
\end{eqnarray}
where
\begin{eqnarray}
  {\mathcal D}_j  
  &=& a^j_4 b^j_2 c^j_5 d^j_1 e^j_3 -  a^j_5 b^j_2 c^j_4 d^j_1 e^j_3
  - a^j_4 b^j_5 c^j_2 d^j_1 e^j_3 +  a^j_5 b^j_4 c^j_2 d^j_1 e^j_3
  \nonumber \\
  &-& a^j_4 b^j_3 c^j_5 d^j_1 e^j_2 +  a^j_5 b^j_3 c^j_4 d^j_1 e^j_2
  + a^j_4 b^j_5 c^j_3 d^j_1 e^j_2 -  a^j_5 b^j_4 c^j_3 d^j_1 e^j_2
  \nonumber \\
  &-& a^j_4 b^j_2 c^j_5 d^j_3 e^j_1 +  a^j_5 b^j_2 c^j_4 d^j_3 e^j_1
  + a^j_4 b^j_5 c^j_2 d^j_3 e^j_1 -  a^j_5 b^j_4 c^j_2 d^j_3 e^j_1
  \nonumber \\
  &+& a^j_4 b^j_3 c^j_5 d^j_2 e^j_1 -  a^j_5 b^j_3 c^j_4 d^j_2 e^j_1
  - a^j_4 b^j_5 c^j_3 d^j_2 e^j_1 +  a^j_5 b^j_4 c^j_3 d^j_2 e^j_1
  \label{eq_A1.16}
\end{eqnarray}

Defining $\epsilon = \sigma_1/\sigma_0$ we may write
\begin{eqnarray}
  {\mathcal D}_j  
  &=& 
    - \epsilon j^3 \alpha_0^{-(2j+1)} \alpha_1^{-(2j+1)}  
    - \epsilon^2 j^2 (j+1) \alpha_0^{-(2j+1)} \alpha_1^{-(2j+1)}   
    - \epsilon j^2 (j+1) \alpha_1^{-(4j+2)}  
    \nonumber \\
    &+& \epsilon^2 j^2 (j+1) \alpha_1^{-(4j+2)}  	
    - j^2 (j+1) \alpha_0^{-(2j+1)} \alpha_1^{-(2j+1)} 
    - \epsilon j (j+1)^2 \alpha_0^{-(2j+1)} \alpha_1^{-(2j+1)}  
    \nonumber \\
    &+& j^2 (j+1) \alpha_1^{-(4j+2)}  
    - \epsilon j^2 (j+1) \alpha_1^{-(4j+2)}  
    - \epsilon j^2 (j+1) \alpha_0^{-(2j+1)}   
    \nonumber \\
    &-&  \epsilon^2 j(j+1)^2 \alpha_0^{-(2j+1)}  
    - \epsilon j(j+1)^2 \alpha_1^{-(2j+1)} 
    + \epsilon^2 j(j+1)^2 \alpha_1^{-(2j+1)}   
    \nonumber \\
    &+& j^2 (j+1) \alpha_0^{-(2j+1)}  
    +  \epsilon j(j+1)^2 \alpha_0^{-(2j+1)} 
    - j^2 (j+1) \alpha_1^{-(2j+1)}  
    \nonumber \\ 
    &+& \epsilon j^2 (j+1) \alpha_1^{-(2j+1)} 
  \label{eq_A1.18b}
\end{eqnarray}

\begin{eqnarray}
  A_{jm} 
  &=& \epsilon (2j+1)^2 \alpha_0^{-(2j+1)} \alpha_1^{-(2j+1)} {\mathcal D}_j^{-1} I_{jm}
  \nonumber \\
  B_{jm} 
  &=& (2j+1)([1 + \epsilon]j + \epsilon) \alpha_0^{-(2j+1)} \alpha_1^{-(2j+1)} {\mathcal D}_j^{-1} I_{jm}
  \nonumber \\
  C_{jm} 
  &=& - (1 - \epsilon)j(2j+1) \alpha_1^{-(2j+1)} {\mathcal D}_j^{-1} I_{jm}
  \nonumber \\
  D_{jm} &=& \left ([(1+\epsilon)j + \epsilon][(1+\epsilon) j + 1] \alpha_0^{-(2j+1)}
  - (1 - \epsilon)^2 j(j+1) \alpha_1^{-(2j+1)} \right) \alpha_1^{-(2j+1)} {\mathcal D}_j^{-1} I_{jm}	
  \nonumber \\
  E_{jm} &=& (1 - \epsilon) j[(1 + \epsilon)j + \epsilon)] 
  [\alpha_0^{-(2j+1)} - \alpha_1^{-(2j+1)}]{\mathcal D}_j^{-1} I_{jm} 
  \label{eq_A1.20h}
\end{eqnarray} 
which can be compactly written as
\begin{eqnarray}
  A_{jm} 
  &=& a_j \alpha_0^{-(2j+1)} \alpha_1^{-(2j+1)} {\mathcal D}_j^{-1} I_{jm}
  \nonumber \\
  B_{jm} 
  &=& b_j \alpha_0^{-(2j+1)} \alpha_1^{-(2j+1)} {\mathcal D}_j^{-1} I_{jm}
  \nonumber \\
  C_{jm} 
  &=& c_j \alpha_1^{-(2j+1)} {\mathcal D}_j^{-1} I_{jm}
  \nonumber \\
  D_{jm} 
  &=& \left[ d_{0j} \alpha_0^{-(2j+1)} + d_{1j} \alpha_1^{-(2j+1)} \right] 
  \alpha_1^{-(2j+1)} {\mathcal D}_j^{-1} I_{jm}  
  \nonumber \\
  E_{jm} 
  &=& e_j \left[\alpha_0^{-(2j+1)} - \alpha_1^{-(2j+1)} \right] {\mathcal D}_j^{-1} I_{jm} 
  \label{eq_A1.20i}
\end{eqnarray} 
where
\begin{eqnarray}
  a_j(\epsilon)	&=& \epsilon (2j+1)^2 
  \nonumber \\
  b_j(\epsilon) &=& (2j+1)([1 + \epsilon]j + \epsilon) 
  \nonumber \\
  c_j(\epsilon) &=& -(1 - \epsilon)j(2j+1)
  \nonumber \\
  d_{0j}(\epsilon) &=& [(1+\epsilon)j + \epsilon][(1+\epsilon) j + 1]   
  \nonumber \\
  d_{1j}(\epsilon) &=& -(1 - \epsilon)^2 j(j+1)   
  \nonumber \\
  e_j(\epsilon) &=& (1-\epsilon) j [(1 + \epsilon) j + \epsilon)]  
  \label{eq_A1.20j}
\end{eqnarray}

\section{Solving for the TMP E Field}
\label{solv_tms}

According to equations (\ref{eq_2.6e})-(\ref{eq_2.6g}) the electric field within $V$ is given by:

\begin{eqnarray}
  {\bf e}_0(r,\theta,\phi,t) &=& - \sum_{jm} [j(2j+1)]^{1/2} A_{jm}(t) r^{j-1} {\bf Y}^{j-1}_{jm}(\theta,\phi)
  \nonumber \\
  &-& \sum_{jm} \frac{r^j}{2j+1} \frac{\partial J^j_{jm}}{\partial t} {\bf Y}^j_{jm}(\theta,\phi)
  - \sum_{jm} \frac{r^{j-1}}{2j-1} \frac{\partial J^{j-1}_{jm}}{\partial t} {\bf Y}^{j-1}_{jm}(\theta,\phi) 	
  \label{app_eq_2.6e}
  \\
  {\bf e}_1(r,\theta,\phi,t) &=&  
  - \sum_{jm} [j(2j+1)]^{1/2} B_{jm}(t) r^{j-1} {\bf Y}^{j-1}_{jm}(\theta,\phi)
  \nonumber \\	
  &-& \sum_{jm} [(j+1)(2j+1)]^{1/2} C_{jm}(t) r^{-(j+2)} {\bf Y}^{j+1}_{jm}(\theta,\phi) 
  \nonumber \\
  &-& \sum_{jm} \frac{r^j}{2j+1} \frac{\partial J^j_{jm}}{\partial t} {\bf Y}^j_{jm}(\theta,\phi)
  - \sum_{jm} \frac{r^{j-1}}{2j-1} \frac{\partial J^{j-1}_{jm}}{\partial t} {\bf Y}^{j-1}_{jm}(\theta,\phi)
  \label{app_eq_2.6f}
  \\
  {\bf e}_2(r,\theta,\phi,t) &=&  
  - \sum_{jm} [j(2j+1)]^{1/2} D_{jm}(t) r^{j-1} {\bf Y}^{j-1}_{jm}(\theta,\phi)
  \nonumber \\	
  &-& \sum_{jm} [(j+1)(2j+1)]^{1/2} E_{jm}(t) r^{-(j+2)} {\bf Y}^{j+1}_{jm}(\theta,\phi) 
  \nonumber \\
  &-& \sum_{jm} \frac{r^j}{2j+1} \frac{\partial J^j_{jm}}{\partial t} {\bf Y}^j_{jm}(\theta,\phi)
  - \sum_{jm} \frac{r^{j-1}}{2j-1} \frac{\partial J^{j-1}_{jm}}{\partial t} {\bf Y}^{j-1}_{jm}(\theta,\phi)
  \label{app_eq_2.6g}
\end{eqnarray}

The boundary conditions are still those given by equations (\ref{eq_1.5})-(\ref{eq_1.6}) albeit with 
${\bf j} \cdot {\bf n} = 0$. Applying the boundary condition at $r=1$ gives:
\begin{eqnarray}
  0
  &=& -\sum_{jm} 
  \left[
    [j(2j+1)]^{1/2} D_{jm} {\bf Y}^{j-1}_{jm}(\theta,\phi)  
    + [(j+1)(2j+1)]^{1/2} E_{jm} {\bf Y}^{j+1}_{jm}(\theta,\phi) 
  \right] \cdot {\hat{\bf r}} 
  \nonumber \\
  &-& 
  \sum_{jm} 
  \left[
    \frac{1}{2j+1} \frac{\partial J^j_{jm}}{\partial t} {\bf Y}^j_{jm}(\theta,\phi)
    + \frac{1}{2j-1} \frac{\partial J^{j-1}_{jm}}{\partial t} {\bf Y}^{j-1}_{jm}(\theta,\phi) 
  \right] \cdot {\hat{\bf r}}
  \nonumber \\
  &=& \sum_{jm} 
  \left[
    j D_{jm} - (j+1) E_{jm} 
    + \left(\frac{j}{2j+1}\right)^{1/2} \frac{1}{2j-1} \frac{\partial J^{j-1}_{jm}}{\partial t}
  \right] Y_{jm}(\theta,\phi)
  \label{eq_A2.71a}
\end{eqnarray}
or
\begin{equation}
  - j D_{jm} + (j+1) E_{jm}  
  = \left( \frac{j}{2j+1}\right)^{1/2} 
  \frac{1}{2j-1}  \frac{\partial J^{j-1}_{jm}}{\partial t}
  \label{eq_A2.71b}
\end{equation}

Applying the first boundary condition of equations (\ref{eq_1.6}) at $r=\alpha_1$ we have
\begin{eqnarray}
  &\sigma_1& \!\!\!\! \sum_{jm} 
  \left[
    [j(2j+1)]^{1/2} B_{jm} \alpha_1^{j-1} {\bf Y}^{j-1}_{jm}(\theta,\phi)  
    + [(j+1)(2j+1)]^{1/2} C_{jm} \alpha_1^{-(j+2)} {\bf Y}^{j+1}_{jm}(\theta,\phi) 
  \right] \cdot {\hat{\bf r}} 
  \nonumber \\
  &+& \sigma_1
  \sum_{jm} 
  \left[
    \frac{\alpha_1^j}{2j+1} \frac{\partial J^j_{jm}}{\partial t} {\bf Y}^j_{jm}(\theta,\phi)
    + \frac{\alpha_1^{j-1}}{2j-1} \frac{\partial J^{j-1}_{jm}}{\partial t} {\bf Y}^{j-1}_{jm}(\theta,\phi) 
  \right] \cdot {\hat{\bf r}}
  \nonumber \\
  &=& 
  \sigma_0 \sum_{jm} 
  \left[
    [j(2j+1)]^{1/2} D_{jm} \alpha_1^{j-1} {\bf Y}^{j-1}_{jm}(\theta,\phi)  
    + [(j+1)(2j+1)]^{1/2} E_{jm} \alpha_1^{-(j+2)} {\bf Y}^{j+1}_{jm}(\theta,\phi) 
  \right] \cdot {\hat{\bf r}} 
  \nonumber \\
  &+& \sigma_0
  \sum_{jm} 
  \left[
    \frac{\alpha_1^j}{2j+1} \frac{\partial J^j_{jm}}{\partial t} {\bf Y}^j_{jm}(\theta,\phi)
    + \frac{\alpha_1^{j-1}}{2j-1} \frac{\partial J^{j-1}_{jm}}{\partial t} {\bf Y}^{j-1}_{jm}(\theta,\phi) 
  \right] \cdot {\hat{\bf r}}	
  \label{eq_A2.7}
\end{eqnarray}
or 
\begin{eqnarray}
  &\sigma_1& \!\!\!\! \sum_{jm} 
  \left[
    j B_{jm} \alpha_1^{j-1} Y_{jm}(\theta,\phi)  
    - (j+1) C_{jm} \alpha_1^{-(j+2)} Y_{jm}(\theta,\phi) 
  \right] 
  \nonumber \\
  &+& (\sigma_1 -\sigma_0)
  \sum_{jm} \left[\frac{j}{2j+1} \right]^{1/2} 
  \frac{\alpha_1^{j-1}}{2j-1} \frac{\partial J^{j-1}_{jm}}{\partial t} Y_{jm}(\theta,\phi) 
  \nonumber \\
  &=& 
  \sigma_0 \sum_{jm} 
  \left[
    j D_{jm} \alpha_1^{j-1} Y_{jm}(\theta,\phi)  
    - (j+1) E_{jm} \alpha_1^{-(j+2)} Y_{jm}(\theta,\phi) 
  \right] 
  \label{eq_A2.72}
\end{eqnarray}
or
\begin{eqnarray}
  & & \sigma_1  
  \left[
    j B_{jm} \alpha_1^{j-1}   
    - (j+1) C_{jm} \alpha_1^{-(j+2)}  
  \right] 
  + (\sigma_1 -\sigma_0)
  \left[\frac{j}{2j+1} \right]^{1/2} 
  \frac{\alpha_1^{j-1}}{2j-1} \frac{\partial J^{j-1}_{jm}}{\partial t}  
  \nonumber \\
  &=& 
  \sigma_0 
  \left[
    j D_{jm} \alpha_1^{j-1}   
    - (j+1) E_{jm} \alpha_1^{-(j+2)}  
  \right] 
  \label{eq_A2.73}
\end{eqnarray}
\begin{eqnarray}
  \epsilon j B_{jm} - \epsilon (j+1) C_{jm}  \alpha_1^{-(2j+1)}  
  - j D_{jm} + (j+1) E_{jm}  \alpha_1^{-(2j+1)} 	
  = (1 - \epsilon)
  \left[\frac{j}{2j+1} \right]^{1/2} 
  \frac{1}{2j-1} \frac{\partial J^{j-1}_{jm}}{\partial t}  
  \label{eq_A2.73a}
\end{eqnarray}
Applying the second boundary condition of equations (\ref{eq_1.6}) at $r=\alpha_1$ we have
\begin{eqnarray}
  & & \sum_{jm} 
  \left[
    [j(2j+1)]^{1/2} B_{jm} \alpha_1^{j-1} {\bf Y}^{j-1}_{jm}(\theta,\phi)  
    + [(j+1)(2j+1)]^{1/2} C_{jm} \alpha_1^{-(j+2)} {\bf Y}^{j+1}_{jm}(\theta,\phi) 
  \right] \times {\hat{\bf r}} 
  \nonumber \\
  &=& 
  \sum_{jm} 
  \left[
    [j(2j+1)]^{1/2} D_{jm} \alpha_1^{j-1} {\bf Y}^{j-1}_{jm}(\theta,\phi)  
    + [(j+1)(2j+1)]^{1/2} E_{jm} \alpha_1^{-(j+2)} {\bf Y}^{j+1}_{jm}(\theta,\phi) 
  \right] \times {\hat{\bf r}} 
  \label{eq_A2.74}
\end{eqnarray}
or
\begin{eqnarray}
  & & \sum_{jm} 
  \left[
    [j(j+1)]^{1/2} B_{jm} \alpha_1^{j-1} {\bf Y}^j_{jm}(\theta,\phi)  
    + [j(j+1)]^{1/2} C_{jm} \alpha_1^{-(j+2)} {\bf Y}^j_{jm}(\theta,\phi) 
  \right] 
  \nonumber \\
  &=& 
  \sum_{jm} 
  \left[
    [j(j+1)]^{1/2} D_{jm} \alpha_1^{j-1} {\bf Y}^j_{jm}(\theta,\phi)  
    + [j(j+1)]^{1/2} E_{jm} \alpha_1^{-(j+2)} {\bf Y}^j_{jm}(\theta,\phi) 
  \right]  
  \label{eq_A2.75}
\end{eqnarray}
or
\begin{eqnarray}
	B_{jm} + C_{jm} \alpha_1^{-(2j+1)} - D_{jm} - E_{jm} \alpha_1^{-(2j+1)} = 0  
  \label{eq_A2.76}
\end{eqnarray}

Applying the first boundary condition of equations (\ref{eq_1.6}) at $r=\alpha_0$ we have
\begin{eqnarray}
  & & \sigma_0 \sum_{jm} 
  \left[
    [j(2j+1)]^{1/2} A_{jm} \alpha_0^{j-1} {\bf Y}^{j-1}_{jm}(\theta,\phi)
    + \frac{\alpha_0^{j-1}}{2j-1}  \frac{\partial J^{j-1}_{jm}}{\partial t} {\bf Y}^{j-1}_{jm}(\theta,\phi) 
  \right] \cdot {\hat{\bf r}} = 
  \nonumber \\
  & & \sigma_1 \sum_{jm} 
  \left[
    [j(2j+1)]^{1/2} B_{jm} \alpha_0^{j-1} {\bf Y}^{j-1}_{jm}(\theta,\phi) 
    + [(j+1)(2j+1)]^{1/2} C_{jm} \alpha_0^{-(j+2)} {\bf Y}^{j+1}_{jm}(\theta,\phi)
  \right] \cdot {\hat{\bf r}}
  \nonumber \\
  &+& \sigma_1
   \frac{\alpha_0^{j-1}}{2j-1}  \frac{\partial J^{j-1}_{jm}}{\partial t} {\bf Y}^{j-1}_{jm}(\theta,\phi) 
   \cdot {\hat{\bf r}} 
\label{eq_A2.8}
\end{eqnarray}
or
\begin{eqnarray}
  & & \sigma_0 \sum_{jm} 
  \left[
    j A_{jm} \alpha_0^{j-1} + \left( \frac{j}{2j+1} \right)^{1/2}
    \frac{\alpha_0^{j-1}}{2j-1} \frac{\partial J^{j-1}_{jm}}{\partial t}  
  \right] Y_{jm}(\theta,\phi) =
  \nonumber \\
  & & \sigma_1 \sum_{jm} 
  \left[
    j B_{jm} \alpha_0^{j-1} - (j+1) C_{jm} \alpha_0^{-(j+2)}
    + \left( \frac{j}{2j+1} \right)^{1/2}
    \frac{\alpha_0^{j-1}}{2j-1}  \frac{\partial J^{j-1}_{jm}}{\partial t}
  \right] Y_{jm}(\theta,\phi)
\label{eq_A2.9}
\end{eqnarray}
or
\begin{eqnarray}
  j A_{jm}
  - \epsilon j B_{jm}  + \epsilon (j+1) C_{jm} \alpha_0^{-(2j+1)}
  =  (\epsilon - 1) \left( \frac{j}{2j+1} \right)^{1/2} \frac{1}{2j-1} 
  \frac{\partial J^{j-1}_{jm}}{\partial t}.
\label{eq_A2.10}
\end{eqnarray}

Applying the second boundary condition of equations (\ref{eq_1.6}) at $r=\alpha_0$ we have
\begin{eqnarray}
  & & \sum_{jm} 
  \left[
    [j(2j+1)]^{1/2} A_{jm} \alpha_0^{j-1} {\bf Y}^{j-1}_{jm}(\theta,\phi)  
  \right] \times {\hat{\bf r}} 
  \nonumber \\
  &=& 
  \sum_{jm} 
  \left[
    [j(2j+1)]^{1/2} B_{jm} \alpha_0^{j-1} {\bf Y}^{j-1}_{jm}(\theta,\phi)  
    + [(j+1)(2j+1)]^{1/2} C_{jm} \alpha_0^{-(j+2)} {\bf Y}^{j+1}_{jm}(\theta,\phi) 
  \right] \times {\hat{\bf r}} 
  \label{eq_A2.77}
\end{eqnarray}
or
\begin{eqnarray}
  & & \sum_{jm} 
  [j(j+1)]^{1/2} A_{jm} \alpha_0^{j-1} {\bf Y}^j_{jm}(\theta,\phi)  
  \nonumber \\
  &=& 
  \sum_{jm} 
  \left[
    [j(j+1)]^{1/2} B_{jm} \alpha_0^{j-1} {\bf Y}^j_{jm}(\theta,\phi)  
    + [j(j+1)]^{1/2} C_{jm} \alpha_0^{-(j+2)} {\bf Y}^j_{jm}(\theta,\phi) 
  \right] 
  \label{eq_A2.78}
\end{eqnarray}
or
\begin{eqnarray}
  A_{jm} - B_{jm} - C_{jm} \alpha_0^{-(2j+1)} = 0
  \label{eq_A2.79}
\end{eqnarray}

We write equations (\ref{eq_A2.72}), (\ref{eq_A2.73a}), (\ref{eq_A2.76}), (\ref{eq_A2.10}) and (\ref{eq_A2.79}) as:
\begin{eqnarray}
  \begin{bmatrix} 
   0 & 0 & 0 & d^{j}_1 &e^{j}_1  \\
   0 & b^{j}_2 & c^{j}_2 & d^{j}_2 & e^{j}_2 \\
   0 & b^{j}_3 & c^{j}_3 & d^{j}_3 & e^{j}_3 \\
   a^{j}_4 & b^{j}_4 & c^{j}_4 & 0 & 0 \\
   a^{j}_5 & b^{j}_5 & c^{j}_5 & 0 & 0
  \end{bmatrix} 
    \begin{bmatrix} 
   A_{jm}  \\
   B_{jm}  \\
   C_{jm}  \\
   D_{jm}  \\
   E_{jm} 
  \end{bmatrix} 
   =  I_{jm}
  \begin{bmatrix} 
   1 \\
   (1 - \epsilon) \\
   0 \\
   (\epsilon - 1) \\
   0
  \end{bmatrix} 
\label{eq_A2.81}
\end{eqnarray}
where
\begin{eqnarray}
 \begin{tabular}{ l l l l l}
   &  &  & $d^{j}_1 = -j $ & $e^{j}_1 = (j+1) $ \\
   & $b^{j}_2 = \epsilon j$ & $c^{j}_2 = -\epsilon (j+1) \alpha_1^{-(2j+1)}$ & $d^{j}_2 = -j$ & $e^{j}_2 = (j+1) \alpha_1^{-(2j+1)}$ \\
   & $b^{j}_3 = 1$ & $c^{j}_3 = \alpha_1^{-(2j+1)} $ & $d^{j}_3 = -1 $ & $e^{j}_3 = -\alpha_1^{-(2j+1)}$ \\
   $a^{j}_4 = j $ & $b^{j}_4 = -\epsilon j$ & $c^{j}_4 = \epsilon (j+1) \alpha_0^{-(2j+1)}$ &  & \\
   $a^{j}_5 = 1$ & $b^{j}_5 = -1$ & $c^{j}_5 = -\alpha_0^{-(2j+1)}$ &  & \\
 \end{tabular}
\label{eq_A2.82}
\end{eqnarray}
and
\begin{eqnarray}
  I_{jm} = \left( \frac{j}{2j+1} \right)^{1/2} \frac{1}{2j-1} \frac{\partial J^{j-1}_{jm}}{\partial t}
\label{eq_A2.82b}
\end{eqnarray}

Using Cramer's Rule the solution of this simultaneous set of equations is:
\begin{eqnarray}
  A_{jm} &=& I_{jm}(d^j_2 e^j_3 - d^j_3 e^j_2)(b^j_4 c^j_5 - b^j_5 c^j_4) {\mathcal D}_j^{-1}
  - I_{jm} (\sigma_2 - \sigma_1) (d^j_1 e^j_3 - d^j_3 e^j_1)(b^j_4 c^j_5 - b^j_5 c^j_4) {\mathcal D}_j^{-1}
  \nonumber \\
  &-& I_{jm} (\sigma_1 - \sigma_0)(d^j_1 e^j_2 - d^j_2 e^j_1)(b^j_3 c^j_5 - b^j_5 c^j_3) {\mathcal D}_j^{-1}
  \nonumber \\	
  &+& I_{jm} (\sigma_1 - \sigma_0)(b^j_2 c^j_5 - b^j_5 c^j_2)(d^j_1 e^j_3 - d^j_3 e^j_1) {\mathcal D}_j^{-1}
  \nonumber \\		
  B_{jm} &=&  I_{jm}(d_3e_2- d_2 e_3)(a_4 c_5 - a_5 c_4) {\mathcal D}_j^{-1}
  +  I_{jm}(\sigma_2 - \sigma_1)(d_1 e_3 - d_3 e_1)(a_4 c_5 - a_5 c_4) {\mathcal D}_j^{-1}
  \nonumber \\
  &+&  I_{jm}(\sigma_1 - \sigma_0) a_5 d_1 (c_2 e_3 - c_3 e_2) {\mathcal D}_j^{-1}
  + I_{jm}(\sigma_1 - \sigma_0) a_5 e_1 (c_3 d_2 - c_2 d_3) {\mathcal D}_j^{-1}
  \nonumber \\	
  C_{jm} &=& I_{jm}(d_2 e_3 - d_3 e_2)(a_4 b_5 - a_5 b_4){\mathcal D}_j^{-1} 
  + I_{jm}(\sigma_2 - \sigma_1)(d_3 e_1 - d_1 e_3)(a_4 b_5 - a_5 b_4) {\mathcal D}_j^{-1}
  \nonumber \\
  &+& I_{jm} a_5 b_2 (\sigma_1 - \sigma_0)(d_3 e_1 - d_1 e_3) {\mathcal D}_j^{-1}
  + I_{jm} a_5 b_3 (\sigma_1 - \sigma_0)(d_1 e_2 - d_2 e_1){\mathcal D}_j^{-1}	
\label{eq_A2.83}
\end{eqnarray}
where
\begin{eqnarray}
  {\mathcal D}_j &=& d_1(b_2 e_3 - b_3 e_2)(a_4 c_5 - a_5 c_4) - e_1(b_2 d_3 - b_3 d_2)(a_4 c_5 - a_5 c_4)
  \nonumber \\
  &-& c_2 (d_1 e_3 - d_3 e_1)(a_4 b_5 - a_5 b_4) + c_3(d_1 e_2 - d_2 e_1)(a_4 b_5 - a_5 b_4) 
\label{eq_A2.84}
\end{eqnarray}
Substituting from equations (\ref{eq_A2.82}) we find
\begin{eqnarray}
  A_{jm} {\mathcal D}_j I_{jm}^{-1} 
  &=& \epsilon (j+1)^2 \alpha_0^{-(2j+1)} \alpha_1^{-(2j+1)} - \epsilon (j+1)^2 \alpha_0^{-(2j+1)} 
  \nonumber \\
  &+& \epsilon^2 j(j+1) \alpha_0^{-(2j+1)} \alpha_1^{-(2j+1)} + \epsilon^2 (j+1)^2 \alpha_0^{-(2j+1)} 	
  \nonumber \\
  &+& \epsilon j(j+1) \alpha_1^{-(4j+2)} - \epsilon j(j+1) \alpha_1^{-(2j+1)} 	
  \nonumber \\
  &-& j(j+1) \alpha_1^{-(4j+2)} + j(j+1) \alpha_0^{-(2j+1)} \alpha_1^{-(2j+1)}	
  \nonumber \\
  &+& j(j+1) \alpha_1^{-(2j+1)} - j(j+1) \alpha_0^{-(2j+1)} 	
  \nonumber \\
  &-& \epsilon^2 j(j+1) \alpha_1^{-(4j+2)} - \epsilon^2 (j+1)^2 \alpha_1^{-(2j+1)} 	
  \nonumber \\
  &+& \epsilon j^2 \alpha_0^{-(2j+1)} \alpha_1^{-(2j+1)} + \epsilon j(j+1) \alpha_0^{-(2j+1)} 	
  \nonumber \\
  &+& \epsilon j(j+1) \alpha_1^{-(4j+2)} + \epsilon (j+1)^2 \alpha_1^{-(2j+1)} 
  \nonumber \\	
  B_{jm} {\mathcal D}_j I_{jm}^{-1} 
  &=& A_{jm} {\mathcal D}_j I_{jm}^{-1} 	  
  \nonumber \\	
  C_{jm} {\mathcal D}_j I_{jm}^{-1} &=& 0
  \nonumber \\	
  D_{jm} {\mathcal D}_j I_{jm}^{-1} &=& A_{jm} {\mathcal D}_j I_{jm}^{-1} 
  \nonumber \\	
  E_{jm} {\mathcal D}_j I_{jm}^{-1} &=& 0
  \label{eq_A2.83b}
\end{eqnarray}
and
\begin{eqnarray}
  {\mathcal D}_j 
  &=& - \epsilon j(j+1)^2 r_0^{-(2j+1)} r_1^{-(2j+1)} + \epsilon j(j+1)^2 r_0^{-(2j+1)} r_2^{-(2j+1)}
  \nonumber \\
  &-& \epsilon^2 j^2(j+1) r_0^{-(2j+1)} r_1^{-(2j+1)} - \epsilon^2 j(j+1)^2 r_0^{-(2j+1)} r_2^{-(2j+1)}	
  \nonumber \\
  &-& \epsilon j^2(j+1) r_1^{-(4j+2)} + \epsilon j^2(j+1) r_1^{-(2j+1)} r_2^{-(2j+1)}	
  \nonumber \\
  &+& j^2(j+1) r_1^{-(4j+2)} - j^2(j+1) r_0^{-(2j+1)} r_1^{-(2j+1)}	
  \nonumber \\
  &-& j^2(j+1) r_1^{-(2j+1)} r_2^{-(2j+1)} + j^2(j+1) r_0^{-(2j+1)} r_2^{-(2j+1)}	
  \nonumber \\
  &+& \epsilon^2 j^2(j+1) r_1^{-(4j+2)} + \epsilon^2 j(j+1)^2 r_1^{-(2j+1)} r_2^{-(2j+1)}	
  \nonumber \\
  &-& \epsilon j^3 r_0^{-(2j+1)} r_1^{-(2j+1)} - \epsilon j^2(j+1) r_0^{-(2j+1)} r_2^{-(2j+1)}	
  \nonumber \\
  &-& \epsilon j^2(j+1) r_1^{-(4j+2)} - \epsilon j(j+1)^2 r_1^{-(2j+1)} r_2^{-(2j+1)}		
  \label{eq_A2.84b}
\end{eqnarray}
Therefore 
\begin{eqnarray}
	A_{jm} = B_{jm} = D_{jm} = -\frac{I_{jm}}{j}  \qquad \qquad C_{jm} = E_{jm} = 0	
  \label{eq_A2.84c}
\end{eqnarray}

By substituting equations (\ref{eq_A2.84c}) into equations (\ref{app_eq_2.6e} - \ref{app_eq_2.6g}) the following 
expression for the electric field in regions 0,1 and 2 is obtained:
\begin{eqnarray}
  {\bf e}(r,\Omega,t) = - \sum_{jm} \frac{r^j}{2j+1} \frac{\partial J^j_{jm}}{\partial t} {\bf Y}^j_{jm}(\theta,\phi)	
  \label{eq_A2.85}
\end{eqnarray}
In fact for an arbitrary number of concentric spherical conductors the expression for ${\bf E}(r,\theta,\phi,t)$ in all
regions will be given by equation (\ref{eq_A2.85}). The electric field due to the surface charge exactly cancels the
$l=j-1$ components of the induced part of the electric field.

\section{Calculating $I_{jm}$ for Typical Electrode Pairs}
\label{I_coef}

Assume there are two TEP electrodes, the first will have outgoing (directed along an outward oriented unit normal vector 
at the surface of the three shell sphere) current and the second will have ingoing current. The center of the first 
electrode is located at the upper pole of the sphere and the center of the second electrode is located at some angle 
$\beta$ relative to the z-axis (the z-axis runs through the poles). The perimeter of each electrode subtends an 
angle $\theta_o$ (from its center) on the surface of the sphere and it is assumed that the radial component of the 
current density provided by the electrodes are uniform and of magnitude $I_o$.  

For the single electrode located at the pole with out-going uniform current density:
\begin{eqnarray}
  I^+_{jm} &=&  I_o \int_0^{\theta_o} \int_0^{2\pi} Y^*_{jm}(\theta,\phi) \sin \theta d\phi d\theta  
  \nonumber \\
  &=& I_o \sqrt{\frac{2j+1}{4\pi} \frac{(j-m)!}{(j+m)!}} \int_0^{2\pi} e^{im\phi} d\phi 
  \int_0^{\theta_o} P_j^m(\cos \theta) \sin \theta d\theta  
  \nonumber \\
  &=&  2\pi \delta_{m0} I_o \sqrt{\frac{2j+1}{4\pi}} 
  \int_1^{\cos \theta_o} P_j(\cos \theta) d\cos \theta  
  \nonumber \\ 
  &=& \delta_{m0} I_o \sqrt{\frac{\pi}{2j+1}}
  [P_{j+1}(\cos \theta) - P_{j-1}(\cos \theta)]_1^{\cos \theta_o}  
  \nonumber \\
  &=& \delta_{m0} I_o \sqrt{\frac{\pi}{2j+1}}
  [P_{j+1}(\cos \theta_o) - P_{j-1}(\cos \theta_o) - P_{j+1}(1) + P_{j-1}(1)]
  \nonumber \\
  &=& \delta_{m0} I_o \sqrt{\frac{\pi}{2j+1}}
  [P_{j+1}(\cos \theta_o) - P_{j-1}(\cos \theta_o)]
  \nonumber \\
  &=& \delta_{m0} 2 \pi I_o \sqrt{\frac{1}{2j+1}}
  \left[ 
    \sqrt{\frac{1}{2j+3}} \tilde{P}_{j+1}(\cos \theta_o) - \sqrt{\frac{1}{2j-1}} \tilde{P}_{j-1}(\cos \theta_o)
  \right]
  \nonumber \\
  &=& \delta_{m0} I^+_{j0}
  \label{eq_ex_3}
\end{eqnarray}
where the $\tilde{P}_j = \sqrt{\frac{2j+1}{4\pi}} P_j$ are the renormalized (numerically stable) Legendre Functions 
and the definition of $I^+_{j0}$ should be obvious.
The $I^+_{jm}$ are by definition the coefficents of a spherical harmonic expansion of the outgoing current 
contribution to the function $I^+(\theta,\phi) = {\bf J}(r_2, \Omega) \cdot {\hat{\bf r}}$ and therefore using 
equation (\ref{eq_ex_3}) we can write
\begin{equation}
  I^+(\theta,\phi) 
  = \sum_j I^+_{j0} \; Y_{j0}(\theta,\phi) = \sum_j \sqrt{\frac{2j+1}{4\pi}} I^+_{j0} \; P_j(\cos \theta)
  \label{eq_ex_4.100}
\end{equation}
The contribution to $I_{jm}$ by a second electrode of the same size rotated to a position $\beta$ relative to 
the pole with ingoing uniform current density can be found by rotating by $\beta$ the function $I^+(\theta,\phi)$ 
for the electrode at the pole given by equation (\ref{eq_ex_4.100}) and changing sign. The $Y_{j0}$ spherical harmonic
transforms under a rotation operator ${\hat D}(\alpha, \beta, \gamma)$ (where $\alpha$, $\beta$ and $\gamma$ are 
Euler angles) according to:
\begin{equation}
  {\hat D}(\alpha, \beta, \gamma) Y_{j0}(\theta,\phi) = 
  \sqrt{\frac{4\pi}{2j+1}} \sum_{m=-j}^{j} Y_{jm}(\theta,\phi)  Y^*_{jm}(\beta,\alpha)
  \label{eq_ex_4b}
\end{equation}
therefore the ingoing contribution to $I^-$ due to the second electrode is
\begin{eqnarray}
  I^-(\Omega) 
  &=& - {\hat D}({\alpha, \beta,\gamma}) I^+(\Omega) 
  \nonumber \\
  &=&  - \sum_j I^+_{j0} {\hat D}({\alpha, \beta,\gamma}) Y_{j0}(\theta,\phi) 
  \nonumber \\
  &=&  -  \sum_j \sum_{m=-j}^{j} I^+_{j0} \sqrt{\frac{4\pi}{2j+1}} Y_{jm}(\theta,\phi)  Y^*_{jm}(\beta,\alpha)
  \label{eq_ex_6}
\end{eqnarray}
and $I^-_{jm}$, the contribution to $I_{jm}$ by the ingoing current density of the second electrode, is given by
\begin{eqnarray}
  I^-_{jm} 
  &=& - \int_1^{-1} \!\!\! \int_0^{2\pi} I^-(\theta,\phi) Y^*_{jm}(\theta, \phi) d\phi \; d\cos \theta
  \nonumber \\
  &=&  - I^+_{j0} \sqrt{\frac{4\pi}{2j+1}} Y^*_{jm}(\beta,\alpha)
  \label{eq_ex_7}
\end{eqnarray}
Here we assume that $\alpha = 0$ and allow $\beta$ to vary the position of the second electrode in which case:
\begin{eqnarray}
  I^-_{jm} \!\!
  &=& - I^+_{j0} \sqrt{\frac{4\pi}{2j+1}} Y^*_{jm}(\beta,0)
  \nonumber \\
  &=& - I^+_{j0} \sqrt{\frac{(j-m)!}{(j+m)!}} P^m_j(\cos \beta)
  \label{eq_ex_8}
\end{eqnarray}
and therefore
\begin{eqnarray}
  I_{jm} &=& I^+_{j0} \left [ \delta_{m0} - \sqrt{\frac{(j-m)!}{(j+m)!}} P^m_j(\cos \beta) \right]
  \nonumber \\
  &=& I^+_{j0} \left [ \delta_{m0} - \sqrt{\frac{4\pi}{2j+1}} {\tilde P}^m_j(\cos \beta) \right]
  \label{eq_ex_9}
\end{eqnarray} 
where the definition of the renormalized associated Legendre functions ${\tilde P}^m_j$ should be obvious.

\section{Calculating $J^j_{jm}$ for Circular and Figure-8 TMS Coils}
\label{J_coef}

The specifications of TMS coils, which contain many windings, are usually given in terms of an inner and outer
radius for a simple circular coil. Here the simple circular coil (see figure \ref{tmp_coil}) is approximated 
by a single winding at the average of the inner and outer radii. Assume the current density $\bf j$ is a thin 
ring of current of amplitude $I(t)$ and radius $r_c$ (in units of $r_2$) inscribed on a plane tangent to the 
outer surface of the scalp region and centered on the vertical axis. Then
\begin{equation}
  {\bf j} = I(t) {\bf e}_{\phi} \delta(r-\rho_o) \delta(\cos \theta - \cos \theta_o) r^{-1} \sin \theta
  \label{eq_ex2_1}
\end{equation}
where $\rho_o = \sqrt{r_c^2 + 1}$ and $\cos \theta_o = \rho_o^{-1}$. It follows that
\begin{eqnarray}
  J^j_{jm} 
  &=& \frac{4\pi r_2^2}{c^2} \iiint \frac{1}{r'^{j+1}} {\bf j}(r', \theta', \phi') \cdot {\bf Y}^{*j}_{jm}(\theta', \phi')
  r'^2 dr' d\phi' d\cos\theta'  
  \nonumber \\
  &=& \frac{4\pi r_2^2}{c^2} I \iiint \frac{1}{r'^{j+1}} 
  \delta(r'- \rho_o) \delta(\cos \theta' - \cos \theta_o) \sin \theta'
  {\bf e}_{\phi} \cdot {\bf Y}^{*j}_{jm}(\theta', \phi') r' dr' d\phi' d\cos\theta'
  \nonumber \\
  &=& \frac{4\pi r_2^2}{c^2 \rho_o^j} I \iint 
  \delta(\cos \theta' - \cos \theta_o) \sin \theta'
  {\bf e}_{\phi} \cdot {\bf Y}^{*j}_{jm}(\theta', \phi') d\phi' d\cos\theta'
  \nonumber \\
  &=& \frac{i}{\sqrt{j(j+1)}} \frac{4\pi r_2^2}{c^2 \rho_o^j} I \iint 
  \delta(\cos \theta' - \cos \theta_o) \sin \theta'
  \frac{\partial Y^*_{jm}}{\partial \theta'} d\phi' d\cos\theta'
  \label{eq_ex2_2}
\end{eqnarray} 
Since 
\begin{equation}
  \frac{\partial Y_{jm}}{\partial \theta'} = \frac{1}{2} \sqrt{j(j+1) - m(m+1)} Y_{jm+1}e^{-i\phi}
  - \frac{1}{2} \sqrt{j(j+1) - m(m-1)} Y_{jm-1}e^{i\phi}
  \label{eq_ex2_3}
\end{equation}
then
\begin{eqnarray}
  J^j_{jm} 
  &=& 
  i \frac{2\pi r_2^2 I}{c^2 \rho_o^j} \left( \frac{j(j+1) - m(m+1)}{j(j+1)} \right)^{1/2} 
  \iint \delta(\cos \theta' - \cos \theta_o) \sin \theta' 
  Y^*_{jm+1}(\theta',\phi') e^{i\phi'} d\phi' d\cos\theta'
  \nonumber \\
  &-&
  i \frac{2\pi r_2^2 I}{c^2 \rho_o^j} \left( \frac{j(j+1) - m(m-1)}{j(j+1)} \right)^{1/2} 
  \iint \delta(\cos \theta' - \cos \theta_o) \sin \theta' 
  Y^*_{jm-1}(\theta',\phi') e^{-i\phi'} d\phi' d\cos\theta'
  \nonumber \\
  &=& 
  i \frac{2\pi r_2^2 I}{c^2 \rho_o^j} \left( \frac{j(j+1) - m(m+1)}{j(j+1)} \right)^{1/2} 
  \iint \delta(\cos \theta' - \cos \theta_o) \sin \theta' 
  e^{-im\phi} {\tilde P}_j^{m+1}(\cos \theta) d\phi' d\cos\theta'
  \nonumber \\
  &-&
  i \frac{2\pi r_2^2 I}{c^2 \rho_o^j} \left( \frac{j(j+1) - m(m-1)}{j(j+1)} \right)^{1/2} 
  \iint \delta(\cos \theta' - \cos \theta_o) \sin \theta' 
  e^{-im\phi} {\tilde P}_j^{m-1}(\cos \theta) d\phi' d\cos\theta'
  \nonumber \\
  &=& 
  i \delta_{m0} \frac{4\pi^2 r_2^2 I}{c^2 \rho_o^j}  
  \int \delta(\cos \theta' - \cos \theta_o) \sin \theta' 
  {\tilde P}_j^{1}(\cos \theta) d\cos\theta'
  \nonumber \\
  &-&
  i \delta_{m0} \frac{4\pi^2 r_2^2 I}{c^2 \rho_o^j}  
  \int \delta(\cos \theta' - \cos \theta_o) \sin \theta' 
  {\tilde P}_j^{-1}(\cos \theta) d\cos\theta'
  \nonumber \\
  &=& 
  i \delta_{m0} \frac{4\pi^2 r_2^2 I}{c^2 \rho_o^j}  
  \sqrt{1 - \cos^2 \theta_o} 
  {\tilde P}_j^{1}(\cos \theta_o) 
  \nonumber \\
  &-&
  i \delta_{m0} \frac{4\pi^2 r_2^2 I}{c^2 \rho_o^j}  
  \sqrt{1 - \cos^2 \theta_o}
  {\tilde P}_j^{-1}(\cos \theta_o) 
  \nonumber \\
  &=& 
  i \delta_{m0} \frac{8\pi^2 r_2^2 I}{c^2 \rho_o^j} \sqrt{1 - \cos^2 \theta_o}
  {\tilde P}_j^{1}(\cos \theta_o) 
  \label{eq_ex2_4}
\end{eqnarray} 
where the identities $Y_{jm}(\theta,\phi) = e^{im\phi} {\tilde P}_j^m(\cos \theta)$ and 
${\tilde P}_j^{-m}(\cos \theta) = (-1)^m {\tilde P}_j^m(\cos \theta)$ have been used.

If a second coil is added with its current circulating in the direction opposite that of coil 1 then a 
figue-of-eight type coil can be obtained. The position and orientation of coil 2 is obtained by rotating coil 
1 by an angle $\beta = 2\theta_o$ from the z-axis such that the two coils osculate (see figure \ref{tmp_coil}) 
at one point. For this figure-8 coil $J^j_{jm} = J^{j+}_{jm} - J^{j-}_{jm}$ where $J^{j+}_{jm}$ is the
contibution from coil 1 (given by equation \ref{eq_ex2_4}) and $J^{j-}_{jm}$ is the contibution from coil 2. 
$J^{j-}_{jm}$ can be found by performing either a rotation of the current density by angle $\beta$ or a 
rotation of the spherical harmonic ${\bf Y}^{*j}_{jm}(\theta, \phi)$ by angle $-\beta$. Using the later approach 
\begin{eqnarray}
  \!\!\!\!\!\!\!\!\!\!\! J^{j-}_{jm} \!\!
  &=& \! \frac{4\pi r_2^2}{c^2} \iiint \frac{1}{r'^{j+1}} {\bf j}(r', \theta', \phi') \cdot 
  \left[ {\hat D}(0,-\beta,0) {\bf Y}^{*j}_{jm}(\theta', \phi') \right]
  r'^2 dr' d\phi' d\cos\theta'  
  \nonumber \\
  &=& \! \frac{4\pi r_2^2}{c^2 \rho_o^j} I 
  \iint \delta(\cos \theta' \! - \cos \theta_o) {\bf e}_{\phi} \cdot
  \left [{\hat D}(0,-\beta,0) {\bf Y}^{*j}_{jm}(\theta', \phi') \right] 
  \sin \theta' d\phi' d\cos\theta'
  \nonumber \\
  &=& \! \frac{4\pi r_2^2}{c^2 \rho_o^j} I 
  \iint \delta(\cos \theta' \! - \cos \theta_o) 
  \sum_{m'} D_{m'm}^j(0,\! -\beta,0) {\bf e}_{\phi} \cdot {\bf Y}^{*j}_{jm'}(\theta', \phi') \sin \theta' d\phi' d\cos\theta'
  \nonumber \\
  &=& \!\! \frac{i 4 \pi I}{\sqrt{j(j+1)}} \frac{r_2^2}{c^2 \rho_o^j} 
  \! \iint \! \delta(\cos \theta' \!- \cos \theta_o) 
  \sum_{m'} D_{m'm}^j(0,\! -\beta,0) \frac{\partial Y^*_{jm'}}{\partial \theta'} \sin \theta' d\phi' d\cos\theta'
  \label{eq_f.22sb3.1}
\end{eqnarray} 
where ${\hat D}(0,-\beta,0)$ is the rotation operator with Euler angle arguments and $ D_{m'm}^j$ are 
the Wigner D-functions \cite{VMK}. But the integration with respect to $\phi'$ yields
\begin{eqnarray}
  \int \frac{\partial Y^*_{jm'}}{\partial \theta'} d\phi' 
  &=& \frac{1}{2} \int \sqrt{j(j+1) - m'(m'+1)} Y^*_{jm'+1}(\theta', \phi') e^{i\phi'} d\phi'
  \nonumber \\
  &-& \frac{1}{2} \int \sqrt{j(j+1) - m'(m'-1)} Y^*_{jm'-1}(\theta', \phi') e^{-i\phi'} d\phi'
  \nonumber \\
  &=& \frac{1}{2} \int \sqrt{j(j+1) - m'(m'+1)} {\tilde P}^{m'+1}_j(\cos \theta') e^{-im'\phi'} d\phi'
  \nonumber \\
  &-& \frac{1}{2} \int \sqrt{j(j+1) - m'(m'-1)} {\tilde P}^{m'-1}_j(\cos \theta') e^{im'\phi'} d\phi'
  \nonumber \\
  &=&  
  \pi \delta_{m'0} \sqrt{j(j+1)}
  \left[ 
    {\tilde P}^{1}_j(\cos \theta') - {\tilde P}^{-1}_j(\cos \theta')
  \right]
  \nonumber \\
  &=&  
  2 \pi \delta_{m'0} \sqrt{j(j+1)} {\tilde P}^{1}_j(\cos \theta') 
  \label{eq_f.22sb3.2}
\end{eqnarray}
Substituting equation (\ref{eq_f.22sb3.2}) into equation (\ref{eq_f.22sb3.1}) and making use of the identity
$D_{0m}^j(\alpha,\beta,\gamma) = \sqrt{4\pi/(2j+1)} Y_{j,-m}(\beta,\gamma)$ \cite{VMK} the result is obtained:
\begin{eqnarray}
  J^{j-}_{jm} 
  &=& i \frac{8\pi^2 r_2^2}{c^2 \rho_o^j} I D_{0m}^j(0,-\beta,0) 
  \int \delta(\cos \theta' - \cos \theta_o) {\tilde P}^{1}_j(\cos \theta')
  \sin \theta' d\cos\theta'
  \nonumber \\
  &=& i \frac{8\pi^2 r_2^2}{c^2 \rho_o^j} I \sqrt{1 - \cos^2 \theta_o} D_{0m}^j(0,-\beta,0) 
  {\tilde P}^{1}_j(\cos \theta_o)
  \nonumber \\
  &=& i \frac{8\pi^2 r_2^2}{c^2 \rho_o^j} I \sqrt{1 - \cos^2 \theta_o} \sqrt{4\pi/(2j+1)} Y_{j,-m}(-\beta,0)
  {\tilde P}^{1}_j(\cos \theta_o)
  \nonumber \\
  &=& i (-1)^m \frac{8\pi^2 r_2^2}{c^2 \rho_o^j} I \sqrt{1 - \cos^2 \theta_o} \sqrt{4\pi/(2j+1)} {\tilde P}_j^m(\cos \beta)
  {\tilde P}^{1}_j(\cos \theta_o)
   \label{eq_f.22sb3.3}
\end{eqnarray} 
which reduces to the result given by equation (\ref{eq_ex2_4}) when $\beta =0$. For the figure-8 coil the 
coefficients $J_{jm}^j$ are then given by:
\begin{eqnarray}
  J^{j}_{jm} 
  &=& J^{j+}_{jm} - J^{j-}_{jm} 
  \nonumber \\
  &=& i \frac{8\pi^2 r_2^2}{c^2 \rho_o^j} I \sqrt{1 - \cos^2 \theta_o} 
  \left[
    \delta_{m0} - (-1)^m \sqrt{\frac{4\pi}{2j+1}} {\tilde P}_j^m(\cos 2\theta_o) 
  \right]
  {\tilde P}^{1}_j(\cos \theta_o)
  \label{eq_f.22sb3.32}
\end{eqnarray}

\bibliography{mag_res}

\end{document}